\documentclass[12pt,oneside]
{article}
\usepackage[latin1]{inputenc}
\usepackage{amsmath}
\usepackage{amsfonts}
\usepackage{amssymb}
\usepackage{graphicx}
\usepackage{authblk}
\usepackage{caption}
\usepackage{subcaption}
\usepackage{sidecap}
\usepackage{color}
\usepackage{chngcntr}
\usepackage{cite}
\counterwithout{figure}{section}

\author[1,2]{Erik D\'iaz-Bautista\footnote{ediaz@fis.cinvestav.mx}}
\author[1]{David J. Fern\'andez C.\footnote{david@fis.cinvestav.mx}}
\affil[1]{\small Physics Department, Cinvestav, P.O. Box 14-740, 07000 Mexico City, Mexico}
\affil[2]{\small Department of Theoretical Physics, Atomic Physics and Optics of the University of Valladolid, 47011 Valladolid, Spain}
\title{Multiphoton supercoherent states}
\date{ }
\begin{document}
	\maketitle

		\begin{abstract}
In this paper we are going to build the multiphoton supercoherent states for the supersymmetric harmonic oscillator as eigenstates of the $m$-th power of a special form (but still with a free parameter) of the Kornbluth-Zypman supersymmetric annihilation operator. They become expressed in terms of the multiphoton coherent states for the standard harmonic oscillator. The Heisenberg uncertainty relation and some statistical properties for these states will be studied. Since the multiphoton supercoherent states turn out to be periodic, then the associated geometric phases will be as well evaluated.
		\end{abstract}

\section{Introduction}
The harmonic oscillator is the simplest exactly solvable quantum mechanical binding model. The ladder operators used to determine the equidistant spectrum for this system generate the well-known Heisenberg-Weyl algebra (HWA), and from them more general algebraic structures can be defined, as the polynomial Heisenberg algebras (PHA) which appear when the standard first-order annihilation and creation operators $\hat{a}$ and $\hat{a}^\dagger$ are replaced by $m$-th order differential ones \cite{fh99,fhn94,fno95,carballo04,bermudez14,celeita16,cdf18}. 

In particular, if the new ladder operators are taken as $\tilde{a}=\hat{a}^m, \ \tilde{a}^\dagger=(\hat{a}^\dagger)^m$, then together with the harmonic oscillator Hamiltonian generate the simplest realization of the PHA. Moreover, in this approach the Hilbert space ${\mathcal H}$ decomposes as the direct sum of $m$ supplementary orthogonal subspaces, ${\mathcal H}= {\mathcal H}_0\oplus \cdots \oplus {\mathcal H}_{m-1}$, with ${\mathcal H}_j$ being generated by the infinite ladder of energy eigenstates which arise when acting iteratively $\tilde{a}^\dagger$ onto the minimum energy eigenstate (the state $\vert j\rangle$ in Fock notation, also called the extremal state of the subspace ${\mathcal H}_j$). If the $m$ infinite ladders of energy eigenvalues are placed together, then the well known harmonic oscillator spectrum is recovered.

Once the PHA for the oscillator has been characterized, it is natural to look for the coherent states as eigenstates of $\tilde{a}$ with complex eigenvalues \cite{gl63a,gl63b,su63,bk95}. It turns out that in each subspace ${\mathcal H}_j$ a family of coherent states exists, which coincide with the so-called multiphoton coherent states (MCS) in the literature \cite{celeita16,cdf18,perelomov72,barut71,buzek90,buzek901,sun92,jex93}. One important property of the MCS is that they are periodic under the evolution induced by $H$, with a period being the fraction $1/m$ of the oscillator period. This indicates that the MCS for $m>1$ are intrinsically quantum cyclic states having associated a geometric phase, which can be calculated explicitly since the general expression for the MCS is known.

On the other hand, the supersymmetric harmonic oscillator (SUSY HO) is a system coming from supersymmetric quantum mechanics (SUSY QM), which combines both bosonic and fermionic oscillators in the supersymmetric Hamiltonian $\hat{H}_{\mathrm{SUSY}}$ as follows \cite{holten,freedman,aragone}:
\begin{equation}\label{n3.2}
	\hat{H}_{\text{SUSY}}= \hat{H}_b - \hat{H}_f,
\end{equation}
with the Hamiltonians for the bosonic and fermionic oscillators being given by
\begin{subequations}
	\begin{eqnarray}\label{n3.3}
		\hat{H}_b & = & \frac{\omega}2 \left\{ \hat{a}^{\dagger}, \hat{a}\right\} = \omega\left(\hat{a}^{\dagger}\hat{a}+\frac12\right) = \omega\left(\hat{N}_b+\frac12\right), \\
		\hat{H}_f & = & \frac{\omega}2 \left[ \hat{f}^{\dagger}, \hat{f}^{-}\right] = \omega\left(\hat{f}^{\dagger}\hat{f}^{-}-\frac12\right) = \omega\left(\hat{N}_f-\frac12\right).
	\end{eqnarray}
\end{subequations}
The bosonic $\hat{a},\hat{a}^{\dagger}$ and fermionic $\hat{f}^{-},\hat{f}^{\dagger}$ annihilation and creation operators satisfy:
\begin{subequations}
	\begin{eqnarray}
		& [\hat{a},\hat{a}^{\dagger}]=1, \qquad \{\hat{f}^{-},\hat{f}^{\dagger}\}=I, \label{n3.4a}\\
		&  \{\hat{f}^{-},\hat{f}^{-}\}=\{\hat{f}^\dagger,\hat{f}^\dagger\}=0,  \label{n3.4b}\\
		& \hat{f}^{-} = f_1 - i f_2, \quad \hat{f}^{\dagger} = f_1 + i f_2, \label{n3.4c}
	\end{eqnarray}
\end{subequations}
where $f_i=\frac12\sigma_i$ with $\sigma_i, i=1,2,3$ being the Pauli matrices, $I$ is the $2\times 2$ identity matrix and $N_b=\hat{a}^{\dagger}\hat{a},$ $N_f=\hat{f}^{\dagger}\hat{f}^{-}$ are the number operators for bosons and fermions respectively. By plugging the expressions (\ref{n3.3}-\ref{n3.4c}) in Eq.~(\ref{n3.2}) it turns out that:
\begin{equation}\label{3.2}
	\hat{H}_{\text{SUSY}}=\omega\left(
	\begin{array}{cc}
		\hat{a}^{\dagger}\hat{a} & 0 \\
		0 & \hat{a}\hat{a}^{\dagger} \\
	\end{array}
	\right),
\end{equation}
whose eigenstates, represented by spinors of two components, and eigenvalues are given by:
\begin{align}
	E_{n}=n\omega, & \quad \vert\Psi^+_n\rangle=\left(
	\begin{array}{c}
		|n\rangle \\
		0 \\
	\end{array}
	\right)=\vert n\rangle\otimes\vert1\rangle_f, \quad \vert\Psi^-_n\rangle=\left(
	\begin{array}{c}
		0 \\
		|n-1\rangle \\
	\end{array}
	\right)=\vert n-1\rangle\otimes\vert0\rangle_f, \label{3.3b}
\end{align}
$n=0,1,\dots$, where $\vert\Psi^-_0\rangle\equiv0$, $\vert1\rangle_f=\left(\!\!
\begin{array}{c c}
1 & 0
\end{array}\!\!
\right)^\mathrm{T}$ and $\vert0\rangle_f=\left(\!\!
\begin{array}{c c}
0 & 1
\end{array}\!\!
\right)^\mathrm{T}$ are fermionic basis vectors. For each natural $n\neq0$, the associated eigenspace is doubly degenerate.

The SUSY HO can be seen as a toy model for the study of the interaction of a cavity mode with a two level system (the matter-radiation interaction). In fact, the SUSY HO Hamiltonian is obtained from a more realistic description of such physical situation supplied by the so-called Jaynes-Cummings model \cite{jc63}, when the oscillating component of the magnetic field is neglected and for exact resonance \cite{dh02}.

The supersymmetric harmonic oscillator Hamiltonian is a matrix which involves two standard harmonic oscillator Hamiltonians (see Eq.~(\ref{3.2})), then it is natural to look for its supersymmetric annihilation operator (SAO) also as a matrix built from the standard annihilation and creation operators. The most obvious SAO were explored some time ago \cite{aragone,hussin}, but a general form including those simple choices was proposed recently by Kornbluth and Zypman as follows \cite{zypman}:
\begin{equation}\label{3.4}
	\hat{A}_{\text{SUSY}}=\left(\begin{array}{cc}
		k_1\hat{a} & k_2 \\
		k_3\hat{a}^2 & k_4\hat{a} \\
	\end{array}\right),
\end{equation}
$k_i\in\mathbb{C}$ being arbitrary parameters. Despite its generality, this form is not unique (as discussed in \cite{aragone,zypman,hussin,erik16}). The previous two proposals are recovered either by taking $k_1=k_4=1$ and $k_2=k_3=0$, which leads to the most obvious diagonal SAO, or through the choice $k_1=k_2=k_4=1$ and $k_3=0$, that produces the simplest non-diagonal SAO \cite{aragone,hussin}.

Let us note that the so-called {\it supercoherent states} $\vert Z\rangle$ are defined as eigenstates of the annihilation operator $\hat{A}_{\text{SUSY}}$ with complex eigenvalues. As usual, they are built as linear combinations of the eigenstates of the SUSY harmonic oscillator in the form
\begin{equation}\label{3.5}
	\vert Z\rangle=\sum\limits_{n=0}^{\infty} a_n \vert\Psi_n^+\rangle + \sum\limits_{n=1}^{\infty} c_n \vert\Psi_n^-\rangle=\sum\limits_{n=0}^{\infty}a_n\left(
	\begin{array}{c}
		\vert n\rangle \\
		0 \\
	\end{array}
	\right)+\sum\limits_{n=1}^{\infty} c_n\left(
	\begin{array}{c}
		0 \\
		\vert n-1\rangle \\
	\end{array}
	\right).
\end{equation}
These states have been studied in some previous works \cite{aragone,zypman,hussin,erik16}, and they can be used also in generalized Jaynes-Cummings models \cite{dd03}, for the construction of superalgebras \cite{k06} and Q-balls \cite{ks98,k98}, among other applications.

Taking into account the previous ideas, our goal in this article is to extend the construction of the multiphoton coherent states for the supersymmetric harmonic oscillator, {\it i.e.}, to build the {\it multiphoton supercoherent states} for the special SAO arising from (\ref{3.4}) by taking $k_1=k_4=1,\ k_3=0$ and to analyze then some of their physical properties. In addition, since the spectrum for this Hamiltonian is equidistant, these states turn out to be periodic, with a period being equal to the fraction $1/m$ of the supersymmetric harmonic oscillator period, thus they have associated geometric phases which can be straightforwardly calculated. 

It is worth to note that the properties of the multiphoton supercoherent states will depend on the parameter $k_2$ of the SAO of Eq.~(\ref{3.4}), which is not fixed. It controls the non-diagonal character of the SAO, since for $\vert k_2\vert \rightarrow 0$ it tends to be diagonal while for large $\vert k_2\vert$ it contains a strong non-diagonal term. We will see that, when $\vert k_2\vert \rightarrow 0$ our multiphoton supercoherent states will have a similar behavior as the multiphoton coherent states for the standard harmonic oscillator while such a behavior will be quite different when $\vert k_2\vert$ is large.

This paper is organized as follows. In sect. \ref{sec2} the polynomial Heisenberg algebras and the multiphoton coherent states for the harmonic oscillator will be quickly reviewed. The Heisenberg uncertainty relation, some statistical and non-classical properties, as well as the periodicity of the MCS, will be also analyzed. In sect. \ref{sec3} the multiphoton supercoherent states will be generated, as eigenstates of the $m$-th power of our special choice of $\hat{A}_{\text{SUSY}}$, and they will be analyzed in the same way as their scalar counterparts of sect. \ref{sec2}. Our conclusions will be presented in sect. \ref{sec4}. The Appendix contains some long expressions which, if they were placed in the body of the text, could compromise the readability of the paper.

\section{Multiphoton coherent states}\label{sec2}
In a quantum mechanical description of the harmonic oscillator the Hamiltonian $\hat{H}$ is usually written as
\begin{equation}\label{2.1}
	\hat{H}=\frac{\hat{p}^2}{2}+\frac{\hat{q}^2}{2},
\end{equation}
where $\hat{q}$ and $\hat{p}$ denote the position and momentum operators, respectively. Defining the ladder operators as
\begin{equation}\label{2.2}
	\hat{a}=\frac{1}{\sqrt{2}}(i\hat{p}+\hat{q}), \quad \hat{a}^\dagger=\frac{1}{\sqrt{2}}(-i\hat{p}+\hat{q}),
\end{equation}
also known as annihilation and creation operators, respectively, it turns out that the set of operators $\{\hat{H},\hat{a},\hat{a}^\dagger\}$ satisfies the commutation relations
\begin{equation}\label{2.4}
	[\hat{H},\hat{a}]=-\hat{a}, \quad [\hat{H},\hat{a}^\dagger]=\hat{a}^\dagger, \quad [\hat{a},\hat{a}^\dagger]=\hat{1},
\end{equation}
which define the so-called Heisenberg-Weyl algebra (HWA).

\subsection{Polynomial Heisenberg algebras}\label{secc2.1}
The previous Heisenberg-Weyl algebra can be deformed by replacing the ladder operators $\hat{a}$ and $\hat{a}^\dagger$ by $m$-th order differential ones $\hat{\mathcal{L}}^-_m$ and $\hat{\mathcal{L}}^\dagger_m$ \cite{fh99,fnd84thesis,dubov92,adler94,sukhatme97,aizawa97,cannata98,arik99,andrianov00}. This leads to the so-called polynomial Heisenberg algebras (PHA) \cite{fh99,fhn94,fno95,carballo04,bermudez14,celeita16}, which are defined by the following commutation relations
\begin{subequations}
	\begin{align}
		&[\hat{H},\hat{\mathcal{L}}^\dagger_m]=\Delta E\,\hat{\mathcal{L}}^\dagger_m\,, \quad [\hat{H},\hat{\mathcal{L}}_m^-]=-\Delta E\,\hat{\mathcal{L}}_m^-\,, \label{4a} \\ &[\hat{\mathcal{L}}_m^-,\hat{\mathcal{L}}^\dagger_m]=\hat{N}_m(\hat{H}+\Delta E\,\hat{1})-\hat{N}_m(\hat{H})\equiv P_{m-1}(\hat{H})\,, \label{4b}
	\end{align}
\end{subequations} 
where the generalized number operator $\hat{N}_m(\hat{H})\equiv\hat{\mathcal{L}}^\dagger_m\hat{\mathcal{L}}^-_m$ is a polynomial in $\hat{H}$ of degree $m$ whose roots are denoted as $\mathcal{E}_i, \ i=1,\dots,m$ so that it can be factorized as
\begin{equation}
	\hat{N}_m(\hat{H})=\prod_{i=1}^{m}(\hat{H}-\mathcal{E}_i),
\end{equation}
$P_{m-1}(\hat{H})$ is a $(m-1)$-th degree polynomial in $\hat{H}$ and $\Delta E>0$ will represent the level spacing inside a given ladder of energy eigenvalues (see below).

\begin{figure}[h!]
	\centering
	\includegraphics[width=1\linewidth]{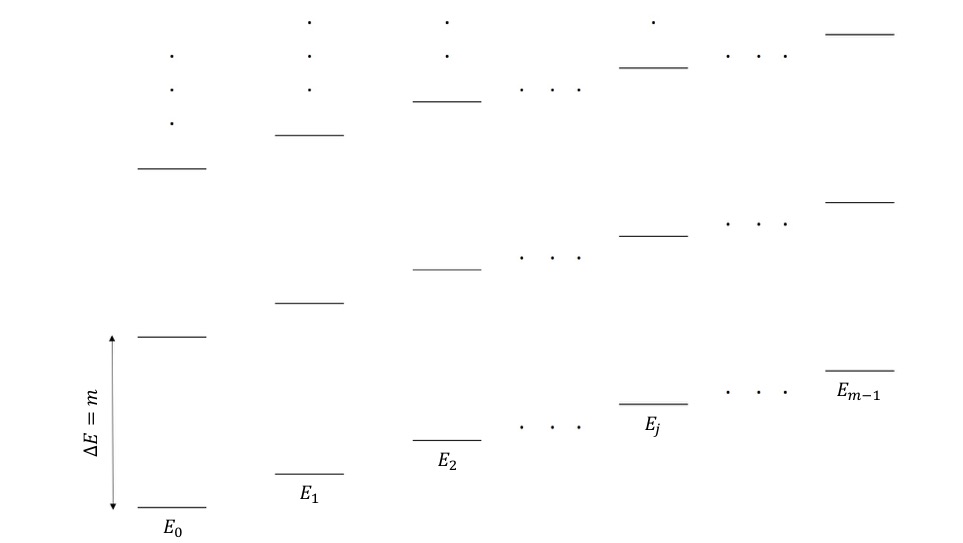}
	\caption[$m$ energy ladders]{The $m$ infinite energy ladders for the harmonic oscillator. The spacing between energy levels in each ladder is $\Delta E=m$.}
	\label{fig:Ladders}
\end{figure}

Let us consider now the set of states $\vert\psi\rangle\in K_{\hat{\mathcal{L}}^-_m}$, where $K_{\hat{\mathcal{L}}^-_m}$ denotes 
the Kernel of the operator $\hat{\mathcal{L}}^-_m$, {\it i.e.},
\begin{equation}
	\hat{\mathcal{L}}^-_m\vert\psi\rangle=0 \quad \Longrightarrow \quad \hat{\mathcal{L}}^\dagger_m\hat{\mathcal{L}}^-_m\vert\psi\rangle=\prod_{i=1}^{m}(\hat{H}-\mathcal{E}_i)\vert\psi\rangle=0.
\end{equation}
Since $K_{\hat{\mathcal{L}}^-_m}$ is invariant under the action of $\hat{H}$, the set of states $\vert\psi_{\mathcal{E}_i}\rangle$ which are simultaneously eigenstates of $\hat{H}$ with eigenvalue $\mathcal{E}_i$ can be selected as the basis of $K_{\hat{\mathcal{L}}^-_m}$, namely,
\begin{equation}
	\hat{\mathcal{L}}^-_m\vert\psi_{\mathcal{E}_i}\rangle=0, \quad
	\hat{H}\vert\psi_{\mathcal{E}_i}\rangle=\mathcal{E}_i\vert\psi_{\mathcal{E}_i}\rangle.
\end{equation}

The $m$ states $\vert\psi_{\mathcal{E}_i}\rangle, \ i=1,\dots,m$ are called extremal states; the remaining eigenstates of $H$ can be constructed by acting $\hat{\mathcal{L}}^\dagger_m$ on $\vert\psi_{\mathcal{E}_i}\rangle$, such that the spacing between energy levels is $\Delta E$. However, if just $s<m$ extremal states are physically meaningful, {\it i.e.}, are eigenstates of $\hat{H}$ satisfying the boundary conditions of the problem, then the iterated action of $\hat{\mathcal{L}}^\dagger_m$ onto $\vert\psi_{\mathcal{E}_i}\rangle$, $i=1,2,\dots,s$ will produce in general a number of energy ladders which is less than the order of the differential operators $\hat{\mathcal{L}}^\pm_m$.

In particular, for the choice \cite{celeita16,cdf18}
\begin{equation}
	\hat{\mathcal{L}}^-_m\equiv\tilde{a}=\hat{a}^m, \quad \hat{\mathcal{L}}^\dagger_m\equiv\tilde{a}^\dagger=\hat{a}^{\dagger m},
\end{equation}
one will be realizing the PHA through the set of operators $\{\hat{H},\tilde{a},\tilde{a}^\dagger\}$, {\it i.e.}, by the harmonic oscillator \cite{dutt99}. In fact, these operators satisfy the algebra defined in the Eqs.~(\ref{4a}, \ref{4b}) with $\Delta E=m$ as follows:
\begin{subequations}
	\begin{align}
		&[\hat{H},\tilde{a}^\dagger]=m\tilde{a}^\dagger\,, \quad [\hat{H},\tilde{a}]=-m\tilde{a}\,, \\ &[\tilde{a},\tilde{a}^\dagger]=\hat{N}_m(\hat{H}+m)-\hat{N}_m(\hat{H})\equiv P_{m-1}(\hat{H})\,, \label{2.14b}
	\end{align}
\end{subequations}
where
\begin{equation}\label{2.15}
	\hat{N}_m(\hat{H}) = \prod_{j=0}^{m-1}(\hat{H}-E_{j}),
\end{equation}
and ${\mathcal{E}}_{j+1}=E_{j}=j+1/2, \ j=0,\dots,m-1$ are the first $m$ eigenvalues of the harmonic oscillator Hamiltonian, that correspond to the $m$ extremal states of the system (see Figure \ref{fig:Ladders}). Thus, in this case the number of energy ladders will coincide with the order of the ladder operators, {\it i.e.}, $s=m$.

The eigenvalues of the $j$-th ladder are
\begin{equation}\label{11}
	E^j_n=E_j+mn, \quad n=0,1,2,\dots, \quad j=0,1,\dots,m-1,
\end{equation}
with the corresponding eigenstates
\begin{equation}
	\vert\psi^j_n\rangle\equiv\vert mn+j\rangle=\sqrt{\frac{j!}{(mn+j)!}}(\tilde{a}^\dagger)^n\vert j\rangle, \quad j=0,1,\dots,m-1.
\end{equation}
Hence, the spectrum of the Hamiltonian $\hat{H}$ becomes:
\begin{equation}
	\text{Sp}(\hat{H})=\{E^0_0,E^0_1,\dots\}\cup\{E^1_0,E^1_1,\dots\}\cup\cdots\cup \{E^j_0,E^j_1,\dots\}\cup\dots\cup\{E^{m-1}_0,E^{m-1}_1,\dots\},
\end{equation}
which is consistent with the standard results $E_n = n+1/2, \ n=0,1,\dots$

We conclude that the Hilbert space $\mathcal{H}=\text{span}(\vert n\rangle,\,n=0,1,2,\ldots)$ has been decomposed as the direct sum of $m$ orthogonal subspaces, $\mathcal{H}=\mathcal{H}_0\oplus\mathcal{H}_2\oplus\cdots	\oplus\mathcal{H}_{m-1}$, where
\begin{equation}\label{2.26}
	\mathcal{H}_{j}=\text{span}(\vert mn+j\rangle,\,n=0,1,2,\ldots), \quad j=0,1,2,\ldots, m-1.
\end{equation}

\subsection{Multiphoton coherent states}\label{secc2.2}
The coherent states were introduced for the first time by Schr\"odinger in 1926 for the harmonic oscillator \cite{schrodinger35}, as wave packets whose dynamics is similar to a classical particle in such a potential. These states have been so widely used in various branches of physics \cite{gl63a,gl63b,su63,bk95,klauder85,zhang90} that nowadays they are called standard coherent states (SCS) in the literature. Moreover, the SCS have been generalized in several different ways \cite{biedenharn89,macfarlane89,klauder63,perelomov86,nieto78}.

One of these generalizations leads to the multiphoton coherent states (MCS), which can be constructed as eigenstates $\vert\tilde{\alpha}\rangle_m$ with complex eigenvalues $\tilde{\alpha}$ of the generalized (multiphoton) annihilation operator $\tilde{a}=\hat{a}^m$ \cite{barut71,buzek90,buzek901,sun92,jex93}:
\begin{equation}\label{2.25}
	\tilde{a}\vert\tilde{\alpha}\rangle_m=
	\tilde{\alpha}\vert\tilde{\alpha}\rangle_m, \quad \tilde{\alpha}\in\mathbb{C}.
\end{equation}

The MCS turn out to be expressed as superpositions of Fock states whose associated energies differ in a quantity which is as a multiple of $m$ (the number of photons required to jump between two levels in the $j$-th energy ladder is a multiple of $m$):
\begin{equation}\label{2.27}
	\vert\tilde{\alpha};j\rangle_m=\mathcal{N}^j_m\sum_{n=0}^{\infty}\frac{\tilde{\alpha}^n}{\sqrt{(mn+j)!}}\vert mn+j\rangle,
\end{equation}
where $\mathcal{N}^j_m$ are normalization constants. The index $j$ indicates that for a given $\tilde{\alpha}$ there is a MCS in each subspace $\mathcal{H}_{j}$ satisfying Eq.~(\ref{2.25}), and the Fock state $\vert j\rangle$ with minimum energy contributing to such a state $\vert\tilde{\alpha};j\rangle_m$ is associated to $E^j_0 = E_j$.

If we choose now $\tilde{\alpha}=\alpha^m$, the MCS take the form
\begin{equation}\label{2.28}
	\vert\alpha;m,j\rangle\equiv\vert\alpha^m;j\rangle_m=\mathcal{N}^j_m\sum_{n=0}^{\infty}\frac{\alpha^{mn+j}}{\sqrt{(mn+j)!}}\vert mn+j\rangle.
\end{equation}
In this approach, the standard coherent states are just a particular case of the multiphoton coherent states for $m=1, \ j=0$:
\begin{equation}\label{2.10}
	\vert\alpha\rangle\equiv\vert\alpha;1,0\rangle=\exp\left(-\frac{\vert\alpha\vert^2}{2}\right)\sum_{n=0}^{\infty}\frac{\alpha^n}{\sqrt{n!}}\vert n\rangle, \quad \tilde{\alpha}=\alpha.
\end{equation}

On the other hand, for $m>1$ the multiphoton coherent states can be also expressed as superpositions of $m$ standard coherent states, which are distributed uniformly on a circle of radius $\vert\alpha\vert$ \cite{clm95,cl12}:
\begin{equation}\label{2.29}
	\vert\alpha;m,j\rangle=\frac{\mathcal{N}^j_m}{m}\sum_{n=0}^{m-1}\omega_j^{\ast n}\left\vert\alpha\omega_j^{n/j}\right\rangle,
\end{equation}
where 
$$
\left\{\omega_j=\exp\left(2\pi i\frac{j}{m}\right);\ j=0,1,\dots,m-1\right\}
$$ 
denotes the abelian group of $m$-th roots of the unit. Such decompositions have been studied in detail in \cite {janszky93,gagen95, jex95,an01, dodonov02,cdf18}.

For example, if we take $m=2$, $j=0,1$ in Eq.~(\ref{2.28}) and the two squared roots of $1$ ($\{\omega_0,\omega_1\}=\{1,e^{i\pi}\}=\{1,-1\}$) in Eq.~(\ref{2.29}), then the explicit forms for the eigenstates of $\tilde{a}=\hat{a}^2$ become:
\begin{subequations}
	\begin{align}
		\vert\alpha\rangle_+\equiv\vert\alpha;2,0\rangle&=\mathcal{N}^0_2\sum_{n=0}^{\infty}\frac{\alpha^{2n}}{\sqrt{(2n)!}}\vert2n\rangle\hspace{1.4cm}=\exp\left(\frac{\vert\alpha\vert^2}{2}\right)\frac{\mathcal{N}^0_2}{2}\left[\vert\alpha\rangle+\vert-\alpha\rangle\right], \label{2.30a} \\
		\vert\alpha\rangle_-\equiv\vert\alpha;2,1\rangle&=\mathcal{N}^1_2\sum_{n=0}^{\infty}\frac{\alpha^{2n+1}}{\sqrt{(2n+1)!}}\vert2n+1\rangle=\exp\left(\frac{\vert\alpha\vert^2}{2}\right)\frac{\mathcal{N}^1_2}{2}\left[\vert\alpha\rangle-\vert-\alpha\rangle\right], \label{2.30b}
	\end{align}
\end{subequations}
where
\begin{equation}
	[\mathcal{N}^0_2]^2=\frac{1}{\cosh(\vert\alpha\vert^2)}, \quad
	[\mathcal{N}^1_2]^2=\frac{1}{\sinh(\vert\alpha\vert^2)}. \label{2.31b}
\end{equation}
This indicates that the states $\vert\alpha\rangle_\pm$ are linear combinations of the two SCS with eigenvalues $\alpha$ and $\alpha e^{i\pi}=-\alpha$. The states in Eqs.~(\ref{2.30a}, \ref{2.30b}) are known either as {\it Schr\"odinger cat states} or {\it even} and {\it odd} coherent states, respectively, since only even (odd) Fock states contribute to the corresponding decomposition \cite{moore96,brune96,haroche,dodonov74,schleich01}.

Similarly, for $m=3$, $j=0,1,2$ and $\tilde{\alpha}=\alpha^3$ we have to consider the three cube roots of $1$, $\{\omega_0,\omega_1,\omega_2\}=\{1,e^{2\pi i/3},e^{4\pi i/3}\}.$ Then, we have
\begin{subequations}
	\begin{align}
		\vert\alpha;3,0\rangle&=\mathcal{N}^0_3\sum_{n=0}^{\infty}\frac{\alpha^{3n}}{\sqrt{(3n)!}}\vert3n\rangle\hspace{1.4cm}=\exp\left(\frac{\vert\alpha\vert^2}{2}\right)\frac{\mathcal{N}^0_3}{3}\left[\vert\alpha\rangle+\vert\alpha \omega_1\rangle+\vert\alpha\omega_2\rangle\right], \label{2.33a}\\
		\vert\alpha;3,1\rangle&=\mathcal{N}^1_3\sum_{n=0}^{\infty}\frac{\alpha^{3n+1}}{\sqrt{(3n+1)!}}\vert3n+1\rangle=\exp\left(\frac{\vert\alpha\vert^2}{2}\right)\frac{\mathcal{N}^1_3}{3}\left[\vert\alpha\rangle+\omega_2\vert\alpha \omega_1\rangle+\omega_1\vert\alpha\omega_2\rangle\right], \label{2.33b}\\
		\vert\alpha;3,2\rangle&=\mathcal{N}^2_3\sum_{n=0}^{\infty}\frac{\alpha^{3n+2}}{\sqrt{(3n+2)!}}\vert3n+2\rangle=\exp\left(\frac{\vert\alpha\vert^2}{2}\right)\frac{\mathcal{N}^2_3}{3}\left[\vert\alpha\rangle+\omega_1\vert\alpha \omega_1\rangle+\omega_2\vert\alpha\omega_2\rangle\right], \label{2.33c}
	\end{align}
\end{subequations}
where
\begin{subequations}
	\begin{eqnarray}
		\mathcal{N}^0_3&=&\left[\frac{1}{3} \left(\exp\left(\vert\alpha\vert^2\right)+2 \exp\left(-\frac{\vert\alpha\vert^2}{2}\right) \cos \left(\frac{\sqrt{3} \vert\alpha\vert^2}{2}\right)\right)\right]^{-1/2}, \label{2.34a}\\
		\mathcal{N}^1_3&=&\left[\frac{1}{3}\left(\exp\left(\vert\alpha\vert^2\right)-2 \exp\left(-\frac{\vert\alpha\vert^2}{2}\right) \sin \left(\frac{\pi}{6} -\frac{\sqrt{3} \vert\alpha\vert^2}{2}\right)\right)\right]^{-1/2}, \label{2.34b} \\
		\mathcal{N}^2_3&=&\left[\frac{1}{3}\left(\exp\left(\vert\alpha\vert^2\right)-2 \exp\left(-\frac{\vert\alpha\vert^2}{2}\right) \sin \left(\frac{\pi}{6}+\frac{\sqrt{3} \vert\alpha\vert^2}{2} \right)\right)\right]^{-1/2}. \label{2.34c}
	\end{eqnarray}
\end{subequations}

As we can see, the states $\vert\alpha,3,j\rangle$ are linear combinations of three SCS with eigenvalues $\alpha$, $\alpha\omega_1$ and $\alpha\omega_2$, which are placed at the vertices of an equilateral triangle in the complex plane $\alpha$. The case with $m=3$ has been considered also in \cite{sun91}, while the multiphoton coherent states for $m=4$ have been addressed in \cite{jex93,hach92,lynch94,moya92,buzek93}.

\subsubsection{Heisenberg uncertainty relation}\label{secc2.2.2}
We can find now joint expressions for the uncertainties associated to the MCS if we define an operator $\hat{s}$ as follows
\begin{equation}
	\hat{s}=\frac{1}{\sqrt{2}i ^k}(\hat{a}+(-1)^k\hat{a}^\dagger), \quad \hat{s}^2=\frac{1}{2}(2\hat{N}+\hat{1}+(-1)^k(\hat{a}^2+\hat{a}^{\dagger 2})), 
\end{equation}
such that
\begin{subequations}
	\begin{align}
		\langle\hat{s}\rangle\vert_{k=0}=\langle\hat{q}\rangle, &\quad \langle\hat{s}\rangle\vert_{k=1}=\langle\hat{p}\rangle, \\
		\langle\hat{s}^2\rangle\vert_{k=0}=\langle\hat{q}^2\rangle, &\quad \langle\hat{s}^2\rangle\vert_{k=1}=\langle\hat{p}^2\rangle,
	\end{align}
\end{subequations}
where $\hat{q}$ and $\hat{p}$ are the position and momentum operators, respectively.

\begin{figure}[h!]
	\centering
	\begin{subfigure}[b]{0.45\textwidth}
		\includegraphics[width=\textwidth]{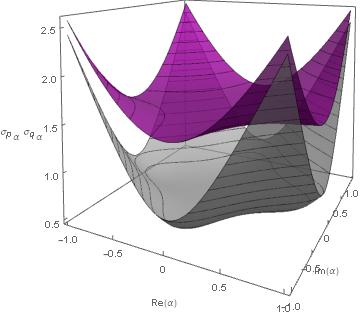}
		\caption{$m=2$, $j=0,1$.}
		\label{fig:Incer_MCS_a}
	\end{subfigure}
	\hspace{1cm}
	~ 
	\begin{subfigure}[b]{0.45\textwidth}
		\includegraphics[width=\textwidth]{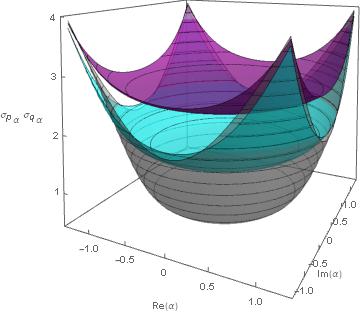}
		\caption{$m=3$, $j=0,1,2$.}
		\label{fig:Incer_MCS_b}
	\end{subfigure}
	\caption{Heisenberg uncertainty relation $\sigma_{q_\alpha}\sigma_{p_\alpha}$ as a function of $\alpha$ for some MCS. This uncertainty takes a minimum value equal to $j+1/2$, $j=0,1,\dots,m-1$, according to the subspace $\mathcal{H}_{j}$ to which the MCS belong.}
	\label{fig:Incer_MCS}
\end{figure}

We thus get:
\begin{subequations}
	\begin{align}
	\langle\hat{s}\rangle_\alpha&=\frac{1}{\sqrt{2}i^k}(\alpha+(-1)^k\alpha^\ast)\delta_{1m}, \\
	\langle\hat{s}^2\rangle_\alpha&=\vert\alpha\vert^2\left[\frac{\mathcal{N}_m^j}{\mathcal{N}_m^{m_j-1}}\right]^2+\frac{1}{2}+(-1)^k([\text{Re}(\alpha)]^2-[\text{Im}(\alpha)]^2)\Delta,
	\end{align}
\end{subequations}
where
\begin{equation}\label{27}
m_j=m\delta_{0j}+j, \qquad \Delta=\begin{cases}
1, & m=1,2, \\
0, & \text{otherwise},
\end{cases} 
\end{equation}
and $\delta_{mn}$ is the Kronecker delta.

In general, the MCS are not minimum uncertainty states, but they fulfill the Heisenberg uncertainty relation (HUR)
\begin{equation}
	\sigma_{q,\alpha}\sigma_{p,\alpha}\geq\frac12.
\end{equation}

Figure \ref{fig:Incer_MCS} shows the HUR for the states $\vert\alpha\rangle_\pm$ and $\vert\alpha;3,j\rangle$, $j=0,1,2$. We can see that in the limit $\alpha\rightarrow0$, this uncertainty achieves its minimum, equal to $j+1/2$, $j=0,1,\dots,m-1$, which coincides with the HUR for the minimum energy eigenstate $\vert j\rangle$ of the harmonic oscillator involved in the linear decomposition for the MCS in the subspace $\mathcal{H}_{j}$ (see Eq.~(\ref{2.28})).

\subsubsection{Non-classicality criteria}\label{secc2.4.1}
There are some criteria in the literature allowing to investigate the non-classical nature of quantum states. In this paper we will focus in the analysis of the sub-Poissonian statistics and the negativity of the Wigner function on phase space, which are useful for such a purpose. In particular, Mandel's $Q$-parameter offers information about the statistics of the states of a quantum system.

\begin{figure}[h!]
	\centering
	\begin{subfigure}[b]{0.450\textwidth}
		\includegraphics[width=\textwidth]{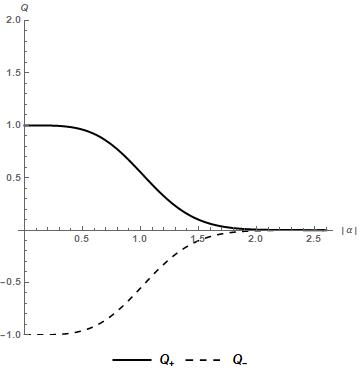}
		\caption{$m=2$, $j=0,1$.}
		\label{fig:mandel_MCS_a}
	\end{subfigure}
	\hspace{1cm}
	~ 
	\begin{subfigure}[b]{0.450\textwidth}
		\includegraphics[width=\textwidth]{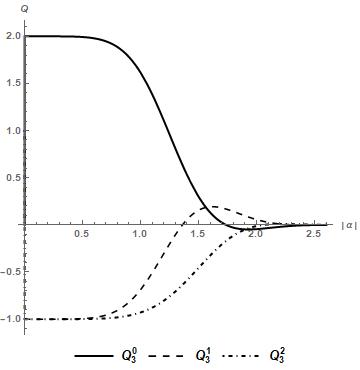}
		\caption{$m=3$, $j=0,1,2$.}
		\label{fig:mandel_MCS_b}
	\end{subfigure}
	\caption{Mandel's $Q$-parameter as function of $|\alpha|$ for some MCS. Poissonian statistics is reached asymptotically for large values of $\vert\alpha\vert$, when the non-classical effects disappear.}
	\label{fig:Mandel_MCS}
\end{figure}

The Mandel's $Q$-parameter is defined as \cite{m79,m82}
\begin{equation}\label{29}
	Q=\frac{\langle\sigma^2_{N}\rangle-\langle\hat{N}\rangle}{\langle\hat{N}\rangle}=\frac{\langle\hat{a}^{\dagger 2}\hat{a}^2\rangle-(\langle\hat{a}^\dagger\hat{a}\rangle)^2}{\langle\hat{a}^\dagger\hat{a}\rangle},
\end{equation}
where $\hat{N}=\hat{a}^\dagger\hat{a}$ is the number operator for the harmonic oscillator. Negative values of $Q$ correspond to {\it quantum states} with a sub-Poissonian statistics, {\it i.e.}, the variance in the photon number is less than its mean. However, if $Q$ takes positive values, no concrete conclusion can be established about the non-classicality of the states. Finally, let us note that the standard coherent states have a Poisson distribution for which $Q=0$.

For the {\it even} and {\it odd} coherent states the Mandel's $Q$-parameter turns out to be (see Figure \ref{fig:mandel_MCS_a})
\begin{subequations}
	\begin{align}
		Q_+&=\vert\alpha\vert^2\left[\coth(\vert\alpha\vert^2)-\tanh(\vert\alpha\vert^2)\right], \\
		Q_-&=\vert\alpha\vert^2\left[\tanh(\vert\alpha\vert^2)-\coth(\vert\alpha\vert^2)\right],
	\end{align}
\end{subequations}
while for the states $\vert\alpha;3,j\rangle$, $j=0,1,2$ it is found that (see Figure \ref{fig:mandel_MCS_b}):
\begin{subequations}
	\begin{align}
		Q_3^0&=\vert\alpha\vert^2\left[\left[\frac{\mathcal{N}^2_3}{\mathcal{N}^{1}_3}\right]^2-\left[\frac{\mathcal{N}^0_3}{\mathcal{N}^{2}_3}\right]^2\right], \\
		Q_3^1&=\vert\alpha\vert^2\left[\left[\frac{\mathcal{N}^0_3}{\mathcal{N}^{2}_3}\right]^2-\left[\frac{\mathcal{N}^1_3}{\mathcal{N}^{0}_3}\right]^2\right], \\
		Q_3^2&=\vert\alpha\vert^2\left[\left[\frac{\mathcal{N}^1_3}{\mathcal{N}^{0}_3}\right]^2-\left[\frac{\mathcal{N}^2_3}{\mathcal{N}^{1}_3}\right]^2\right].
	\end{align}
\end{subequations}
Figure \ref{fig:Mandel_MCS} shows also that the Mandel's $Q$-parameter tends asymptotically to zero when $\vert\alpha\vert\rightarrow\infty$ in both cases \cite{erhg13}.

\begin{figure}[h!]
	\centering
	\begin{subfigure}[b]{0.45\textwidth}
		\includegraphics[width=\textwidth]{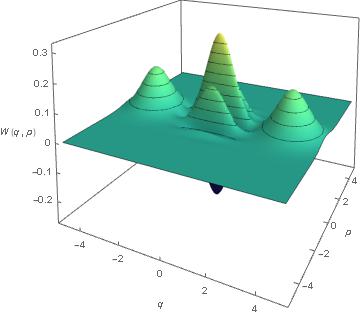}
		\caption{$W^+_\alpha(q,p)$}
		\label{fig:Wigner_20}
	\end{subfigure}
	\hspace{1cm}
	~ 
	\begin{subfigure}[b]{0.45\textwidth}
		\includegraphics[width=\textwidth]{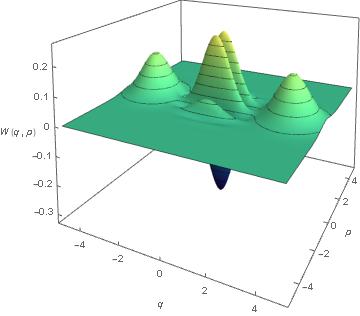}
		\caption{$W^-_\alpha(q,p)$}
		\label{fig:Wigner_21}
	\end{subfigure}
	\caption{Wigner function $W_{\alpha}(q,p)$ for the {\it even} (a) and {\it odd}  (b) coherent states with $\vert\alpha\vert=2.5$.}\label{fig:Wigner_2}
\end{figure}

\begin{figure}[h!]
	\centering
	\begin{subfigure}[b]{0.45\textwidth}
		\includegraphics[width=\textwidth]{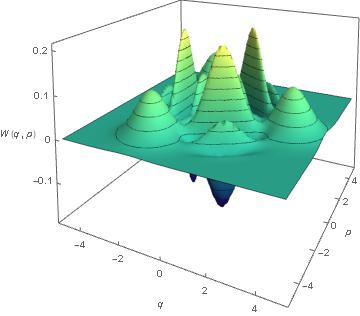}
		\caption{$W^0_{\alpha,3}(q,p)$}
		\label{fig:Wigner_30}
	\end{subfigure}
	\hspace{1cm}
	~ 
	\begin{subfigure}[b]{0.45\textwidth}
		\includegraphics[width=\textwidth]{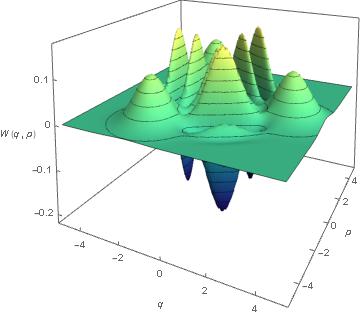}
		\caption{$W^1_{\alpha,3}(q,p)$}
		\label{fig:Wigner_31}
	\end{subfigure}
	\bigskip
	~ 
	\begin{subfigure}[b]{0.45\textwidth}
		\includegraphics[width=\textwidth]{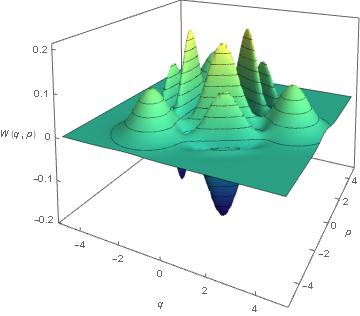}
		\caption{$W^2_{\alpha,3}(q,p)$}
		\label{fig:Wigner_32}
	\end{subfigure}
	\caption{Wigner function $W_{\alpha}(q,p)$ for the MCS $\vert\alpha;3,0\rangle$ (a), $\vert\alpha;3,1\rangle$ (b) and $\vert\alpha;3,2\rangle$ (c) with $\vert\alpha\vert=2.5$.}\label{fig:Wigner_3}
\end{figure}

On the other hand, the Wigner function on phase space is defined as \cite{wigner32,kenfack04,hillery84,cahill96}
\begin{equation}\label{2.58}
	W(q,p)\equiv\frac{1}{2\pi\hbar}\int_{-\infty}^{\infty}\left\langle q-\frac{y}{2}\right\vert\hat{\rho}\left\vert q+\frac{y}{2}\right\rangle \exp\left(\frac{i}{\hbar}py\right)dy,
\end{equation}
where $\hat{\rho}$ is the density operator and $\vert q\pm y\rangle$ are eigenkets of the position operator $\hat{q}$. If the state under analysis is pure, then $\hat{\rho}=\vert \psi\rangle\langle\psi\vert$ and hence (by taking $\hbar=1$):
\begin{equation}
	W_\psi(q,p)=\frac{1}{\pi}\int_{-\infty}^{\infty}\psi^\ast(q+y)\psi(q-y)\exp\left(2ipy\right)dy, \label{2.59}
\end{equation}
with $\langle q-y\vert\psi\rangle=\psi(q-y)$. Negative values arising in the Wigner function indicate the non-classicality of a state, which is interpreted as a sign of {\it quantumness} \cite{kenfack04,smithey93,dunn95,breitenbach97,banaszek99,lougovski03,lamb69,wodkiewicz84,royer85,banaszek96}.

For the standard coherent state $\vert\alpha\rangle$ of Eq.~(\ref{2.10}) and its corresponding wavefunction in coordinates representation \cite{dodonov02}, it is straightforward to find the Wigner function for two SCS with different complex labels $\alpha, \, \beta$:
\begin{equation}\label{2.63}
	W_{\alpha,\beta}(q,p)=\frac{1}{\pi}\exp\left(-\left[q-\frac{(\beta+\alpha^\ast)}{\sqrt{2}}\right]^2-\left[p-\frac{(\beta-\alpha^\ast)}{\sqrt{2}i}\right]^2+\alpha^\ast\beta-\frac{1}{2}\left[\vert\alpha\vert^2+\vert\beta\vert^2\right] \right).
\end{equation}
In particular, $W_{\alpha}(q,p)$ appears from the previous expression for $\beta=\alpha$.

By using now Eq.~(\ref{2.63}), it is possible to find compact expressions for the Wigner functions of the {\it even} and {\it odd} coherent states, respectively (see Figure \ref{fig:Wigner_2}):
\begin{subequations}
	\begin{align}
		W_{\alpha}^+(q,p)&=\exp\left(\vert\alpha\vert^2\right)\frac{[\mathcal{N}_2^0]^2}{4}\left[W_{\alpha}(q,p)+ W_{-\alpha}(q,p)+2\mathrm{Re}[W_{\alpha,-\alpha}(q,p)]\right], \label{2.66a} \\
		W_{\alpha}^-(q,p)&=\exp\left(\vert\alpha\vert^2\right) \frac{[\mathcal{N}_2^1]^2}{4}\left[W_{\alpha}(q,p)+W_{-\alpha}(q,p)-2\mathrm{Re}[W_{\alpha,-\alpha}(q,p)]\right]. \label{2.66b} 
	\end{align}
\end{subequations}
The first two terms in Eqs.~(\ref{2.66a}, \ref{2.66b}) correspond to the Gaussian functions for the two standard coherent states centered in $\pm (q_0,p_0)$. The last is an interference term that oscillates quickly as the distance between the two standard coherent states grows \cite{leonhardt97}. These oscillating terms induce negative values in the Wigner function, thus the {\it even} and {\it odd} coherent states are non-classical states \cite{braverman12}.

On the other hand, Eq.~(\ref{2.63}) allows also to find simply the Wigner function for the MCS with $m=3$, $j=0,1,2$ (see Figure \ref{fig:Wigner_3}):
\begin{subequations}
	\begin{align}
		\nonumber	W^0_{\alpha,3}(q,p)&=\exp\left(\vert\alpha\vert^2\right)\frac{[\mathcal{N}_3^0]^2}{9}\left[W_{\alpha}(q,p)+W_{\alpha\omega_1}(q,p)+W_{\alpha\omega_2}(q,p)\right. \\
		&\quad\left.+2\mathrm{Re}[W_{\alpha,\alpha\omega_1}(q,p)+W_{\alpha,\alpha\omega_2}(q,p)+W_{\alpha\omega_1,\alpha\omega_2}(q,p)]\right], \label{2.68a} \\
		\nonumber	W^1_{\alpha,3}(q,p)&=\exp\left(\vert\alpha\vert^2\right)\frac{[\mathcal{N}_3^1]^2}{9}\left[W_{\alpha}(q,p)+W_{\alpha\omega_1}(q,p)+W_{\alpha\omega_2}(q,p)\right.  \\
		&\quad\left.+2\mathrm{Re}[\omega_2W_{\alpha,\alpha\omega_1}(q,p)+\omega_1W_{\alpha,\alpha\omega_2}(q,p)+\omega^{\ast}_1W_{\alpha\omega_1,\alpha\omega_2}(q,p)]\right], \label{2.68b} \\
		\nonumber	W^2_{\alpha,3}(q,p)&=\exp\left(\vert\alpha\vert^2\right)\frac{[\mathcal{N}_3^2]^2}{9}\left[W_{\alpha}(q,p)+W_{\alpha\omega_1}(q,p)+W_{\alpha\omega_2}(q,p)\right. \\
		&\quad\left.+2\mathrm{Re}[\omega_1W_{\alpha,\alpha\omega_1}(q,p)+\omega_2W_{\alpha,\alpha\omega_2}(q,p)+\omega_1W_{\alpha\omega_1,\alpha\omega_2}(q,p)]\right]. \label{2.68c} 
	\end{align}
\end{subequations}
As for the {\it even} and {\it odd} coherent states, Eqs.~(\ref{2.68a}-\ref{2.68c}) include once again interference terms that contribute mainly in the zone between the Gaussian functions of the three standard coherent states, the last ones being placed in the vertices of an equilateral triangle in phase space. Due to the interference terms, negative values for the Wigner function appear, thus the states in Eqs.~(\ref{2.33a}-\ref{2.33c}) are also non-classical states.

\subsubsection{Evolution loop}\label{secc2.5.3}
The dynamics of a quantum system is determined by its evolution operator, a unitary operator $\hat{U}(t)$ which satisfies
\begin{equation}\label{2.89}
	\frac{d\hat{U}(t)}{dt}=-i\hat{H}(t)\hat{U}(t), \quad \hat{U}(0)=\hat{1},
\end{equation}
where $\hat{H}(t)$ is the system Hamiltonian and $\hat{1}$ represents the identity operator. 

Of special interest in this approach are the so-called evolution loops (EL), {\it i.e.}, dynamical processes such that $\hat{U}(t)$ becomes the identity operator (up to a phase factor) at a certain time $\tau>0$:
\begin{equation}\label{2.90}
	\hat{U}(\tau)=\exp(i\phi)\hat{1},
\end{equation}
where $\tau$ is the loop period and $\phi\in\mathbb{R}$ \cite{mielnik77,mielnik86,fnd94,fndS12}. The EL, introduced by Mielnik in 1977 \cite{mielnik77,mielnik86}, are important since they can be used as the basis to implement control techniques and to perform dynamical manipulation of quantum systems.

If a system performs an evolution loop and an arbitrary state $\vert\psi\rangle\in\mathcal{H}$ is taken as an initial condition, $\vert\psi(0)\rangle\equiv \vert\psi\rangle$, then a cyclic state of period $\tau$ is produced, namely:
\begin{equation}\label{2.91}
	\vert\psi(\tau)\rangle=\hat{U}(\tau)\vert\psi\rangle=\exp(i\phi)\vert\psi\rangle.
\end{equation}
The total phase $\phi$ has a geometric component $\beta$, which in general is not zero. If the system's Hamiltonian is time-independent, such a geometric phase becomes \cite{fnd94}
\begin{equation}\label{2.93}
	\beta=\phi+\frac{1}{\hbar}\int_{0}^{\tau}\langle\psi(0)\vert\hat{U}^\dagger(t)\hat{H}\hat{U}(t)\vert\psi(0)\rangle dt=\phi+\frac{\tau}{\hbar}\langle\psi\vert\hat{H}\vert\psi\rangle.
\end{equation}

In particular, for the harmonic oscillator it is produced an evolution loop of period $T=2\pi$ in the Hilbert space $\mathcal{H}$. Moreover, when $\mathcal{H}$ is decomposed as in section~\ref{secc2.1}, $\mathcal{H} = \mathcal{H}_0\oplus\dots\oplus \mathcal{H}_{m-1}$, it turns out that in each subspace a {\it partial evolution loop} of period $\tau=2\pi/m$ appears, which implies that any state in $\mathcal{H}_j$ is cyclic with period $\tau=2\pi/m$. As the MCS $\vert \alpha;m,j\rangle$ belongs to $\mathcal{H}_j$, then it has also such a period and thus a geometric phase $\beta_j$ is induced, which is given by:
\begin{equation}
	\beta_{j}=-\frac{2\pi }{m}E^j_0+\frac{2\pi}{m}\left[\vert\alpha\vert^2\left[\frac{\mathcal{N}^j_m}{\mathcal{N}^{m_j-1}_m}\right]^2+\frac{1}{2}\right], \quad j=0,1,\dots,m-1. \label{2.105b}
\end{equation}
This result has been recently obtained in \cite{cdf18}.

\section{Multiphoton supercoherent states}\label{sec3}
We are going to build now the coherent states $|Z\rangle_m$ for the supersymmetric harmonic oscillator as eigenstates of the $m$-th power of the SAO. For simplicity, let us choose the parameters of Eq.~(\ref{3.4}) as $k_1=k_4=1$, $k_3=0$ while $k_2\in\mathbb{R}$ is left arbitrary, so that the multiphoton supersymmetric annihilation operator to be used is:
\begin{equation}\label{3.14}
	\hat{A}^m_{\text{SUSY}}=\left(\begin{array}{cc}
		\hat{a}^m & mk_2\hat{a}^{m-1} \\
		0 & \hat{a}^m \\
	\end{array}\right)=\hat{a}^m\otimes I+mk_2\hat{a}^{m-1}\otimes \hat{f}^\dagger.
\end{equation}
This choice will allow us to analyze the effects of the free parameter $k_2$ on the properties of the multiphoton supercoherent states. Before doing this, however, let us sketch first the result of acting the supersymmetric ladder operators $\hat{A}^m_{\text{SUSY}}, \ \hat{A}^{\dagger m}_{\text{SUSY}}$ onto the system's Hilbert space.

As in the previous section, the Hilbert space $\mathcal{H}_{\text{SUSY}}=\text{span}(\vert\Psi_n^+\rangle,\vert\Psi_n^-\rangle,\, n=0,1,\ldots)$ is again decomposed as the direct sum of $m$ orthogonal subspaces $\mathcal{H}_{\text{SUSY}}=\mathcal{H}_0\oplus\mathcal{H}_1\oplus\cdots\oplus\mathcal{H}_{m-1}$, where each subspace
\begin{equation}\label{3.17}
	\mathcal{H}_{j}=\text{span}(\vert\Psi_{mn+j}^+\rangle,\vert\Psi_{mn+j}^-\rangle,\, n=0,1,2,\ldots), \quad j=0,1,2,\ldots, m-1,
\end{equation}
is invariant under the action of $\hat{A}^m_{\text{SUSY}}, \ \hat{A}^{\dagger m}_{\text{SUSY}}$. Meanwhile, the spectrum of the Hamiltonian $\hat{H}_{\text{SUSY}}$ is expressed as
\begin{equation}
	\text{Sp}(\hat{H}_\text{SUSY})=\{E^0_0,E^0_1,\dots\}\cup\{E^1_0,E^1_1,\dots\}\cup\dots\cup\{E^{m-1}_0,E^{m-1}_1,\dots\},
\end{equation}
whose spacing between neighbor energy levels $E^j_n=mn+E_j$ in each subspace $\mathcal{H}_{j}$ is $\Delta E^j_n=m$.

Now, taking into account the coherent states definition,
\begin{equation}\label{3.15}
	\hat{A}^m_{\text{SUSY}}\vert Z\rangle_m=\alpha\vert Z\rangle_m, \quad \alpha\in\mathbb{C},
\end{equation}
and using the expansion in Eq.~(\ref{3.5}), the following relations for the coefficients $a_n$ and $c_n$ are obtained:
\begin{subequations}
	\begin{eqnarray}
		c_{mn+m_j}&=&\sqrt{\frac{(m_j-1)!}{(mn+m_j-1)!}}\alpha^nc_{m_j} , \quad j=0,1,\ldots,m-1, \label{3.16a} \\
		a_{mn+j}&=&\sqrt{\frac{j!}{(mn+j)!}}\alpha^na_j-\sqrt{\frac{(m_j-1)!}{(mn+j)!}}\,mn\,k_2\alpha^{n-\delta_{0j}}c_{m_j}, \quad j=0,1,\ldots,m-1, \label{3.16b}
	\end{eqnarray}
\end{subequations}
where $a_j$ and $c_{m_j}$ are $2m$ free parameters.

If we make now $\alpha=z^m$ in Eqs.~(\ref{3.16a}, \ref{3.16b}), the multiphoton supercoherent states become:
\begin{equation}\label{3.19}
	\vert Z;m,j\rangle=\tilde{a}_{m_j}\vert Z;m,j\rangle_f+\tilde{c}_{m_j} \vert \tilde{Z};m,j\rangle_s,
\end{equation}
where
\begin{equation}\label{59}
	\vert Z;m,j\rangle_f=\left(
	\begin{array}{c}
		\vert z;m,j\rangle \\
		0 \\
	\end{array}
	\right), \quad \vert \tilde{Z};m,j\rangle_s=\left(
	\begin{array}{c}
		-k_2\vert z';m,j\rangle \\
		\vert z;m,m_j-1\rangle \\
	\end{array}
	\right),
\end{equation}
with
\begin{subequations}
	\begin{eqnarray}
		\tilde{a}_{m_j}&=&\sqrt{j!}z^{-j}a_j+jk_2z^{-1}\tilde{c}_{m_j}, \label{3.20a} \\
		\tilde{c}_{m_j}&=&\sqrt{(m_j-1)!}z^{-(m_j-1)}c_{m_j}, \label{3.20b}\\
		\vert z;m,j\rangle&=&
		\sum_{n=0}^{\infty}\frac{z^{mn+j}}{\sqrt{(mn+j)!}}\vert mn+j\rangle, \label{3.20c} \\
		\vert z';m,j\rangle&=&\frac{d}{dz}\vert z;m,j\rangle=\hat{a}^\dagger\vert z;m,m_j-1\rangle. \label{3.20d}
	\end{eqnarray}
\end{subequations}
The states in Eqs.~(\ref{3.20c}, \ref{3.20d}) are not normalized and $m_j$ is again given by Eq.~(\ref{27}). By taking now $m=1$, $j=0$ and $k_2=0$, the supercoherent states for the simplest diagonal SAO are recovered \cite{hussin,zypman}, while the results for a nondiagonal SAO, which mixes both bosonic and fermionic components, are obtained for $m=1$, $j=0$ and $k_2=1$ \cite{aragone}.

According to Eq.~(\ref{59}), the multiphoton supercoherent states $\vert Z;m,j\rangle$ in general are expressed in terms of (scalar) multiphoton coherent states $\vert z;m,j\rangle$ which belong to different Hilbert subspaces $\mathcal{H}_{j}$. It is possible to introduce a new set of states $\vert Z;m,j\rangle_s$ belonging to the supercoherent two-dimensional subspace, Eq.~(\ref{3.19}). Since they are also eigenstates of the operator $\hat{A}^m_{\text{SUSY}}$, without loss of generality we can take as multiphoton supercoherent states those given by:
\begin{equation}\label{55}
	\vert \widetilde{Z;m,j}\rangle=\chi_1\vert Z;m,j\rangle_f+\chi_2\vert Z;m,j\rangle_s, \quad \chi_1,\chi_2\in\mathbb{C},
\end{equation}
where
\begin{equation}
	\vert Z;m,j\rangle_s=\frac{1}{\sqrt{2}}\left(
	\begin{array}{c}
		k_2z^\ast\vert z;m,s_j\rangle-k_2\vert z';m,s_j\rangle \\
		\vert z;m,j\rangle \\
	\end{array}
	\right), 
\end{equation}
with
\begin{equation}
	\vert z';m,s_j\rangle=\hat{a}^\dagger\vert z;m,j\rangle, \quad s_j=(j+1)(1-\delta_{(m-1)j}).
\end{equation}

The multiphoton supercoherent states of Eq.~(\ref{55}) with $m=1$, $j=0$ and $k_2=1$ have been studied extensively in \cite{aragone}, while those with $m=2$, $j=0,1$ have been built recently as {\it even} and {\it odd} superpositions of supercoherent states \cite{amaj16}.

In the next sections, we will analyze only the multiphoton supercoherent states $\vert Z;m,j\rangle$ of Eq.~(\ref{3.19}), which were obtained directly as eigenstates of the annihilation operator $\hat{A}_{\text{SUSY}}^m$.

\subsection{Heisenberg uncertainty relation}\label{secc3.2.1}
In order to analyze the HUR for the multiphoton supercoherent states, let us consider an extension of the operator $\hat{s}$ as $\hat{s}\rightarrow\hat{s}\otimes I$, {\it i.e.}, we take the following matrix operator:
\begin{equation}
	\hat{s}=\frac{1}{\sqrt{2}i^k}\left(\begin{array}{c c}
		\hat{a}+(-1)^k\hat{a}^\dagger & 0 \\
		0 & \hat{a}+(-1)^k\hat{a}^\dagger
	\end{array}\right). \label{3.1.58}
\end{equation}

For the multiphoton supercoherent states $\vert Z;m,j\rangle$ of Eq.~(\ref{3.19}), the mean value of the observable $\hat{s}$ is
\begin{equation}\label{3.21}
	\langle\hat{s}\rangle=\frac{\langle Z;m,j\vert\hat{s}\vert Z;m,j\rangle}{\langle Z;m,j\vert Z;m,j\rangle},
\end{equation}
where
\begin{align}\label{3.22}
	&\nonumber\langle Z;m,j\vert\hat{s}\vert Z;m,j\rangle= \vert\tilde{a}_{m_j}\vert^2\langle z;m,j\vert\hat{s}\vert z;m,j\rangle+\vert\tilde{c}_{m_j}\vert^2\langle z;m,m_j-1\vert\hat{s}\vert z;m,m_j-1\rangle\\
	&+\vert\tilde{c}_{m_j}\vert^2k_2^2\langle z';m,j\vert\hat{s}\vert z';m,j\rangle-k_2\left(\tilde{a}_{m_j}\tilde{c}^\ast_{m_j}\langle z';m,j\vert\hat{s}\vert z;m,j\rangle+\tilde{a}^\ast_{m_j}\tilde{c}_{m_j}\langle z;m,j\vert\hat{s}\vert z';m,j\rangle\right).
\end{align}

\subsubsection{Heisenberg uncertainty relation for $m=1$.}
First of all, let us express the multiphoton supercoherent states with $m=1,\ j=0$ in terms of the normalized standard coherent states $\vert z\rangle$ of Eq.~(\ref{2.10}):
\begin{equation}\label{61}
	\vert Z\rangle\equiv\vert Z;1,0\rangle=\mathcal{N}\exp\left(\frac{\vert z\vert^2}{2}\right)\left[a_0\left(\begin{array}{c}
		\vert z\rangle \\
		0
	\end{array}\right)+c_1\left(\begin{array}{c}
		-k_2\sqrt{\vert z\vert^2+1}\,\hat{a}^\dagger\vert z\rangle \\
		\vert z\rangle
	\end{array}\right)\right],
\end{equation}
where $\mathcal{N}$ is the normalization constant given by
\begin{equation}
	\mathcal{N}^2=\exp(-\vert z\vert^2)\left[\vert a_0\vert^2+\vert c_1\vert^2+\vert c_1\vert^2k_2^2(\vert z\vert^2+1)-2k_2\mathrm{Re}[a_0^\ast c_1z^\ast]\right]^{-1} .
\end{equation}

By taking now $m=1,\ j=0$ in Eq.~(\ref{3.21}) it is found an explicit expression for $(\sigma_s)^2_Z$ (see Eq.~(\ref{sm1j0}) in the Appendix). Let us note that $(\sigma_q)^2_Z=(\sigma_s)^2_Z\vert_{k=0}$ and $(\sigma_p)^2_Z=(\sigma_s)^2_Z\vert_{k=1}$. 

\begin{figure}[h!]
	\centering
	\begin{subfigure}[b]{0.45\textwidth}
		\includegraphics[width=\textwidth]{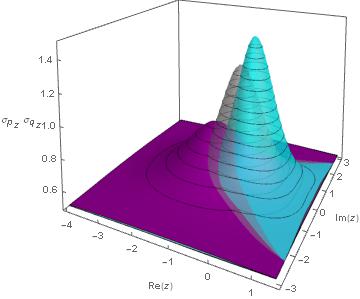}
		\caption{$k_2=-0.8$ (purple), $k_2=-2$ (gray) and $k_2=-10$ (cyan).}
		\label{fig:Inck2SAO_k_1_0a}
	\end{subfigure}
	\hspace{1cm}
	~ 
	\begin{subfigure}[b]{0.45\textwidth}
		\includegraphics[width=\textwidth]{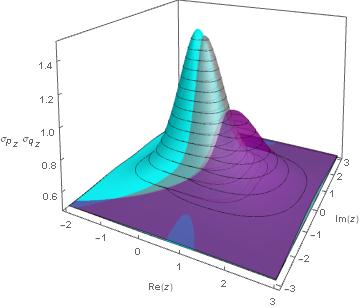}
		\caption{$k_2=1$ (purple), $k_2=5$ (gray) and $k_2=50$ (cyan).}
		\label{fig:Inck2SAO_k_1_0b}
	\end{subfigure}
	\caption{\label{fig:Inck2SAO_k_1_0}Heisenberg uncertainty relation $(\sigma_q)_Z(\sigma_p)_Z$ as function of $z$ for the states $\vert Z\rangle$ with $a_0=c_1=1$ and different values of $k_2$.}
\end{figure}

For $k_2=0$ the Heisenberg uncertainty relation as function of $z$ for the states $\vert Z\rangle$ can be drawn as a constant plane, since then $(\sigma_q)_Z(\sigma_p)_Z=1/2$. However, for $k_2\neq 0$ it appears a maximum for this uncertainty, which tends to the limit value $\sim1.5$ as $\vert k_2\vert $ grows (see Figure \ref{fig:Inck2SAO_k_1_0}); on the other hand, when $\vert z\vert\rightarrow\infty$ the HUR decreases quickly, approaching asymptotically its lowest possible value ($1/2$).

\subsubsection{Heisenberg uncertainty relation for $m=2$.}
Let us express now the multiphoton supercoherent states with $m=2$, $j=0,1$ in terms of the normalized {\it even} and {\it odd} coherent states $\vert z\rangle_\pm$ of Eqs.~(\ref{2.30a}, \ref{2.30b}):
\begin{subequations}\label{64}
	\begin{align}
		\vert Z\rangle_+&\equiv\vert Z;2,0\rangle=\mathcal{N}_+\sqrt{\cosh\left(\vert z\vert^2\right)}\left[\tilde{a}_{2_0}\left(\begin{array}{c}
			\vert z\rangle_+ \\
			0
		\end{array}\right)+\tilde{c}_{2_0}\left(\begin{array}{c}
			-k_2\sqrt{\vert z\vert^2+\tanh(\vert z\vert^2)}\,\hat{a}^\dagger\vert z\rangle_- \\
			\sqrt{\tanh(\vert z\vert^2)}\vert z\rangle_-
		\end{array}\right)\right], \label{64a} \\
		\vert Z\rangle_-&\equiv\vert Z;2,1\rangle=\mathcal{N}_-\sqrt{\sinh\left(\vert z\vert^2\right)}\left[\tilde{a}_{2_1}\left(\begin{array}{c}
			\vert z\rangle_- \\
			0
		\end{array}\right)+\tilde{c}_{2_1}\left(\begin{array}{c}
			-k_2\sqrt{\vert z\vert^2+\coth(\vert z\vert^2)}\,\hat{a}^\dagger\vert z\rangle_+ \\
			\sqrt{\coth(\vert z\vert^2)}\vert z\rangle_+
		\end{array}\right)\right], \label{64b}
	\end{align}
\end{subequations}
where the normalization constantes $\mathcal{N}_\pm$ are given by

\begin{figure}[h!]
	\centering
	\begin{subfigure}[b]{0.45\textwidth}
		\includegraphics[width=\textwidth]{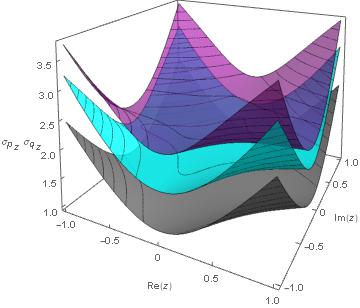}
		\caption{$k_2=-2$ (purple), $k_2=0$ (gray) and $k_2=1$ (cyan).}
		\label{fig:Inck2SAO_k_2_0a}
	\end{subfigure}
	\hspace{1cm}
	~ 
	\begin{subfigure}[b]{0.45\textwidth}
		\includegraphics[width=\textwidth]{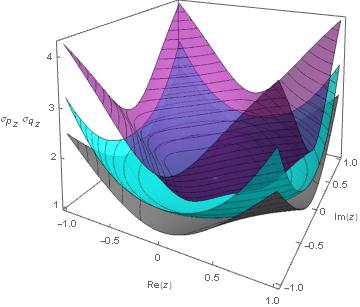}
		\caption{$k_2=-2$ (purple), $k_2=0$ (gray) and $k_2=1$ (cyan).}
		\label{fig:Inck2SAO_k_2_0b}
	\end{subfigure}
	\caption{\label{fig:Inck2SAO_k_2_0}Heisenberg uncertainty relation $(\sigma_q)_Z(\sigma_p)_Z$ as function of $z$ for the states $\vert Z\rangle_+$ with $a_0=c_2=1$ (a), and $\vert Z\rangle_-$ with $a_1=c_1=1$ (b), for different values of $k_2$.}
\end{figure}

\begin{subequations}
	\begin{align}
		&\mathcal{N}_+^2=\frac{\left[\vert \tilde{a}_{2_0}\vert^2+\vert \tilde{c}_{2_0}\vert^2\tanh(\vert z\vert^2)+\vert \tilde{c}_{2_0}\vert^2k_2^2(\vert z\vert^2+\tanh(\vert z\vert^2))-2k_2\tanh(\vert z\vert^2)\mathrm{Re}[\tilde{a}_{2_0}^\ast \tilde{c}_{2_0}z^\ast]\right]^{-1}}{\cosh(\vert z\vert^2)}, \\
		&\mathcal{N}_-^2=\frac{\left[\vert \tilde{a}_{2_1}\vert^2+\vert \tilde{c}_{2_1}\vert^2\coth(\vert z\vert^2)+\vert \tilde{c}_{2_1}\vert^2k_2^2(\vert z\vert^2+\coth(\vert z\vert^2))-2k_2\coth(\vert z\vert^2)\mathrm{Re}[\tilde{a}_{2_1}^\ast \tilde{c}_{2_1}z^\ast]\right]^{-1}}{\sinh(\vert z\vert^2)},
	\end{align}
\end{subequations}
with
\begin{subequations}
	\begin{eqnarray}
		\tilde{a}_{2_0}=a_0, &\quad& \tilde{c}_{2_0}=z^{-1}c_2, \\
		\tilde{a}_{2_1}=z^{-1}a_1+k_2z^{-1}c_1, &\quad& \tilde{c}_{2_1}=c_1.
	\end{eqnarray}
\end{subequations}

By taking now $m=2$, $j=0,1$ in Eq.~(\ref{3.21}) we obtain the explicit expressions for $(\sigma_s)^2_{Z+}$ and $(\sigma_s)^2_{Z-}$ given in equations (\ref{3.28a}) and (\ref{3.28b}) in the Appendix. Let us note once again that $(\sigma_q)^2_{\pm Z}=(\sigma_s)^2_{\pm Z}\vert_{k=0}$ and $(\sigma_p)^2_{\pm Z}=(\sigma_s)^2_{\pm Z}\vert_{k=1}$.

Figures \ref{fig:Inck2SAO_k_2_0a} and \ref{fig:Inck2SAO_k_2_0b} show the Heisenberg uncertainty relation as function of $z$ for $\vert Z\rangle_+$ and $\vert Z\rangle_-$, respectively, and several values of the parameter $k_2$. As we can observe, such a behavior is similar to that obtained for the {\it cat states} (see Figure \ref{fig:Incer_MCS}a). It can be seen also that the minimum value that $(\sigma_q)_Z(\sigma_p)_Z$ can take for both states $\vert Z\rangle_\pm$ grows as $\vert k_2\vert$ does.

\begin{figure}[h!]
	\centering
	\begin{subfigure}[b]{0.45\textwidth}
		\includegraphics[width=\textwidth]{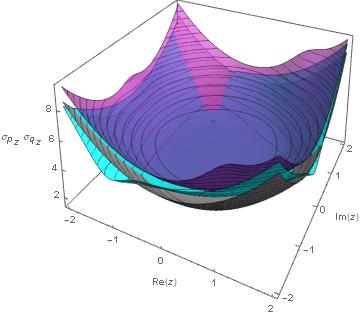}
		\caption{$k_2=-2$ (purple), $k_2=0$ (gray) and $k_2=1$ (cyan).}
		\label{fig:Inck2SAO_k_3_0a}
	\end{subfigure}
	\hspace{1cm}
	~ 
	\begin{subfigure}[b]{0.45\textwidth}
		\includegraphics[width=\textwidth]{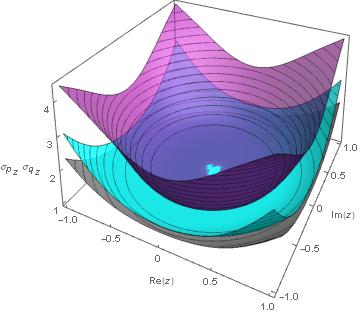}
		\caption{$k_2=-2$ (purple), $k_2=0$ (gray) and $k_2=1$ (cyan).}
		\label{fig:Inck2SAO_k_3_0b}
	\end{subfigure}
	\bigskip
	~ 
	\begin{subfigure}[b]{0.45\textwidth}
		\includegraphics[width=\textwidth]{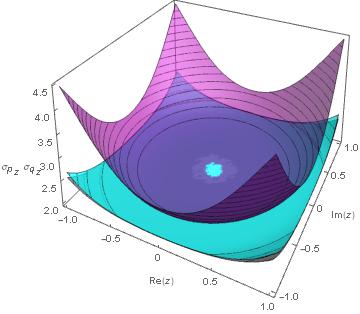}
		\caption{$k_2=-2$ (purple), $k_2=0$ (gray) and $k_2=1$ (cyan).}
		\label{fig:Inck2SAO_k_3_0c}
	\end{subfigure}
	\caption{\label{fig:Inck2SAO_k_3_0}Heisenberg uncertainty relation $(\sigma_q)_Z(\sigma_p)_Z$ as function of $z$ for the states $\vert Z;3,0\rangle$ with $a_0=c_3=1$ (a), $\vert Z;3,1\rangle$ with $a_1=c_1=1$ (b), and $\vert Z;3,2\rangle$ with $a_2=c_2=1$ (c), for different values of $k_2$.}
\end{figure}

\subsubsection{Heisenberg uncertainty relation for $m=3$.}
Finally, for $m=3$, $j=0,1,2$, we get the following multiphoton supercoherent states expressed in terms of the normalized multiphoton coherent states $\vert z;3,j\rangle$ of Eqs.~(\ref{2.33a}-\ref{2.33c}):
\begin{subequations}\label{68}
	\begin{align}
		\vert Z;3,0\rangle&=\mathcal{N}_0[\mathcal{N}_3^0]^{-1}\left[\tilde{a}_{3_0}\left(\begin{array}{c}
			\vert z;3,0\rangle \\
			0
		\end{array}\right)+\tilde{c}_{3_0}\left(\begin{array}{c}
			-k_2\mathcal{N}_3^0\hat{a}^\dagger\vert z;3,2\rangle \\
			\mathcal{N}_3^0[\mathcal{N}_3^{2}]^{-1}\vert z;3,2\rangle
		\end{array}\right)\right], \label{68a}\\
		\vert Z;3,1\rangle&=\mathcal{N}_1[\mathcal{N}_3^1]^{-1}\left[\tilde{a}_{3_1}\left(\begin{array}{c}
			\vert z;3,1\rangle \\
			0
		\end{array}\right)+\tilde{c}_{3_1}\left(\begin{array}{c}
			-k_2\mathcal{N}_3^1\hat{a}^\dagger\vert z;3,0\rangle \\
			\mathcal{N}_3^1[\mathcal{N}_3^{0}]^{-1}\vert z;3,0\rangle
		\end{array}\right)\right], \label{68b}\\
		\vert Z;3,2\rangle&=\mathcal{N}_2[\mathcal{N}_3^2]^{-1}\left[\tilde{a}_{3_2}\left(\begin{array}{c}
			\vert z;3,2\rangle \\
			0
		\end{array}\right)+\tilde{c}_{3_2}\left(\begin{array}{c}
			-k_2\mathcal{N}_3^2\hat{a}^\dagger\vert z;3,1\rangle \\
			\mathcal{N}_3^2[\mathcal{N}_3^{1}]^{-1}\vert z;3,1\rangle
		\end{array}\right)\right], \label{68c}
	\end{align}
\end{subequations}
where the normalization constants $\mathcal{N}_j$ are
\begin{subequations}\label{69}
	\begin{align}
		&[\mathcal{N}_0]^2=\left[\vert\tilde{a}_{3_0}\vert^2[\mathcal{N}_3^0]^{-2}+\vert\tilde{c}_{3_0}\vert^2[\mathcal{N}^{2}_3]^{-2}+\vert\tilde{c}_{3_0}\vert^2k_2^2(\vert z\vert^2[\mathcal{N}^{1}_3]^{-2}+[\mathcal{N}^{2}_3]^{-2})-2k_2[\mathcal{N}^{2}_3]^{-2}\mathrm{Re}[\tilde{a}_{3_0}^\ast\tilde{c}_{3_0}z^\ast]\right]^{-1}, \\
		&[\mathcal{N}_1]^2=\left[\vert\tilde{a}_{3_1}\vert^2[\mathcal{N}_3^1]^{-2}+\vert\tilde{c}_{3_1}\vert^2[\mathcal{N}^{0}_3]^{-2}+\vert\tilde{c}_{3_1}\vert^2k_2^2(\vert z\vert^2[\mathcal{N}^{2}_3]^{-2}+[\mathcal{N}^{0}_3]^{-2})-2k_2[\mathcal{N}^{0}_3]^{-2}\mathrm{Re}[\tilde{a}_{3_1}^\ast\tilde{c}_{3_1}z^\ast]\right]^{-1}, \\
		&[\mathcal{N}_2]^2=\left[\vert\tilde{a}_{3_2}\vert^2[\mathcal{N}_3^2]^{-2}+\vert\tilde{c}_{3_2}\vert^2[\mathcal{N}^{1}_3]^{-2}+\vert\tilde{c}_{3_2}\vert^2k_2^2(\vert z\vert^2[\mathcal{N}^{0}_3]^{-2}+[\mathcal{N}^{1}_3]^{-2})-2k_2[\mathcal{N}^{1}_3]^{-2}\mathrm{Re}[\tilde{a}_{3_2}^\ast\tilde{c}_{3_2}z^\ast]\right]^{-1},
	\end{align}
\end{subequations}
with the constants $\mathcal{N}_3^j$, $j=0,1,2$ given by Eqs.~(\ref{2.34a}-\ref{2.34c}) and
\begin{subequations}
	\begin{eqnarray}
		\tilde{a}_{3_0}=a_0, &\quad& \tilde{c}_{3_0}=\sqrt{2!}z^{-2}c_3, \\
		\tilde{a}_{3_1}=z^{-1}a_1+k_2z^{-1}c_1, &\quad& \tilde{c}_{3_1}=c_1, \\
		\tilde{a}_{3_2}=\sqrt{2!}z^{-2}a_2+2k_2z^{-2}c_2, &\quad& \tilde{c}_{3_2}=z^{-1}c_2.
	\end{eqnarray}
\end{subequations}

If we take now $m=3$, $j=0,1,2$ in Eq.~(\ref{3.21}) we will get the uncertainty relation associated to the operator $\hat{s}$ for this case (see Eq.~(\ref{3.31}) in the Appendix). It is clear once again that $(\sigma_q)^2_{jZ}=(\sigma_s)^2_{jZ}\vert_{k=0}$, $(\sigma_p)^2_{jZ}=(\sigma_s)^2_{jZ}\vert_{k=1}$, $j=0,1,2$.

Figure \ref{fig:Inck2SAO_k_3_0} shows that, in general, the Heisenberg uncertainty relation as function of $z$ for the states $\vert Z;3,j\rangle$ behaves qualitatively in the same way as the one obtained for the scalar states $\vert\alpha;3,j\rangle$ (compare Figure \ref{fig:Incer_MCS_b}). 

Finally, according to the previous discussions, the case of $k_2=0$ constitutes a limit situation for the HUR, since the corresponding gray surfaces are below the other ones for most of the considered cases. This means that, in general, although the multiphoton supercoherent states are not minimum uncertainty states, their HUR reach the allowed minimum values for $k_2=0$.


\subsection{Non-classicality criteria}\label{secc3.2.2}
In order to analyze in more detail the quantumness of the multiphoton supercoherent states, we are going to consider next the Mandel's $Q$-parameter and the Wigner function $W_{Z,m}(q,p)$.

\begin{figure}[h!]
	\centering
	\begin{subfigure}[b]{0.4\textwidth}
		\includegraphics[width=\textwidth]{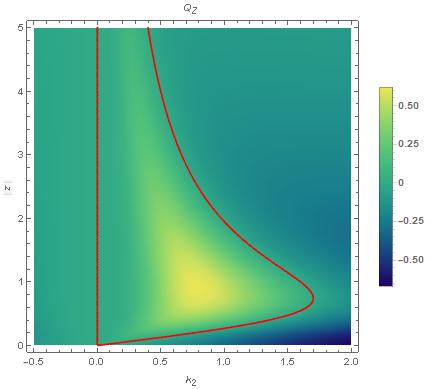}
		\caption{$z\in\mathbb{R}$}
		\label{fig:Wigk2SAO_k_11a}
	\end{subfigure}
	\hspace{1cm}
	~ 
	\begin{subfigure}[b]{0.4\textwidth}
		\includegraphics[width=\textwidth]{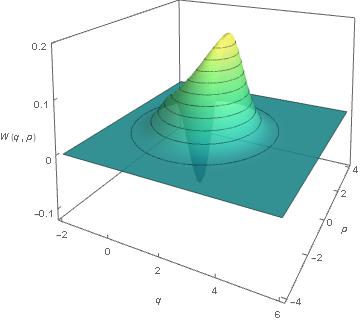}
		\caption{$\vert z\vert=1$, $k_2=1.6$}
		\label{fig:Wigk2SAO_k_11b}
	\end{subfigure}
	\bigskip
	~ 
	\begin{subfigure}[b]{0.4\textwidth}
		\includegraphics[width=\textwidth]{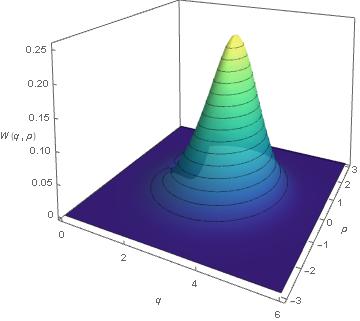}
		\caption{$\vert z\vert=2$, $k_2=0.97561$}
		\label{fig:Wigk2SAO_k_11c}
	\end{subfigure}
	\hspace{1cm}
	~ 
	\begin{subfigure}[b]{0.4\textwidth}
		\includegraphics[width=\textwidth]{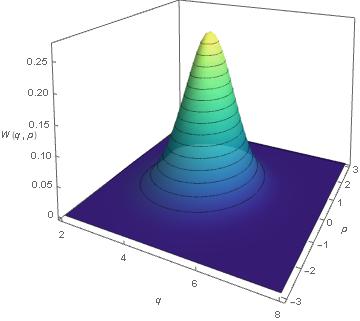}
		\caption{$\vert z\vert=3$, $k_2=0.66298$}
		\label{fig:Wigk2SAO_k_11d}
	\end{subfigure}
	\caption{\label{fig:Wigk2SAO_k_11}Mandel's $Q_Z$-parameter (a) and  Wigner function $W_{Z}(q,p)$ (b-d) for the states in Eq.~(\ref{61}) with $a_0=c_1=1$ and different values of $\vert z\vert, \ k_2$. The red line for $Q_Z$ marks the values of $k_2$ for which a Poissonian statistics is observed.}
\end{figure}

For the multiphoton supercoherent states of Eq.~(\ref{3.19}), the $Q$-parameter is given by Eq.~(\ref{29}), while the corresponding Wigner function $W_{Z,m}(q,p)$ is expressed as
\small
\begin{align}\label{3.35}
	\nonumber   W^j_{Z,m}(q,p)&=\frac{1}{\pi}\int_{-\infty}^{\infty}\Psi^{\dagger j}_{Z,m}(q+y)\Psi^j_{Z,m}(q-y) \exp\left(2ipy\right)dy=|\tilde{a}_{m_j}|^2W^j_{z,m}(q,p)+|\tilde{c}_{m_j}|^2W^{m_j-1}_{z,m}(q,p) \\
	&\quad+|\tilde{c}_{m_j}|^2k_2^2W'^j_{z,m}(q,p)-k_2\left(\tilde{a}^\ast_{m_j}\tilde{c}_{m_j}W^{jII}_{z,m}(q,p)+\tilde{a}_{m_j}\tilde{c}_{m_j}^\ast W^{jI}_{z,m}(q,p)\right),
\end{align}
\normalsize
where $\Psi^j_{Z,m}(q)\equiv\langle q\vert Z;m,j\rangle$, $W^j_{z,m}(q,p)$ denotes the Wigner function for the scalar wavefunction $\psi^j_{z,m}(q)\equiv\langle q\vert z;m,j\rangle$, $W'^j_{z,m}(q,p)$ represents the Wigner function for the scalar wavefunction $\psi'^j_{z,m}(q)=\langle q\vert z';m,j\rangle$, which is given by
\begin{align}
	\nonumber W^{'}_{\alpha,\beta}(q,p)&=\frac{1}{\pi}\int_{-\infty}^{\infty}\psi'^{\ast}_{\alpha}(q+y)\psi'_{\beta}(q-y)\exp\left(2ipy\right)dy\\
	&=2\left(\left[q-\frac{(\beta+\alpha^\ast)}{\sqrt{2}}\right]^2+\left[p-\frac{(\beta-\alpha^\ast)}{\sqrt{2}i}\right]^2-\frac{1}{2}\right)W_{\alpha,\beta}(q,p),
\end{align}

\begin{figure}[h!]
	\centering
	\begin{subfigure}[b]{0.4\textwidth}
		\includegraphics[width=\textwidth]{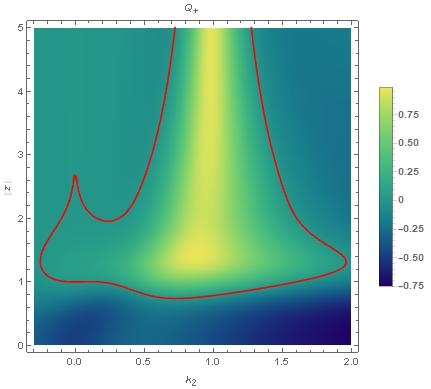}
		\caption{$z\in\mathbb{R}$}
		\label{fig:Wigk2SAO_k_2_0a}
	\end{subfigure}
	\hspace{1cm}
	~ 
	\begin{subfigure}[b]{0.4\textwidth}
		\includegraphics[width=\textwidth]{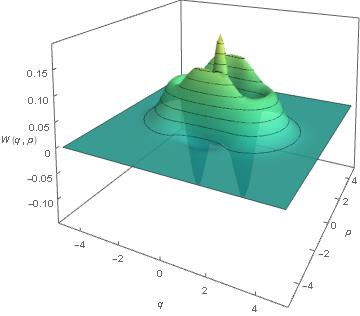}
		\caption{$\vert z\vert=1$, $k_2=1.598698$}
		\label{fig:Wigk2SAO_k_2_0b}
	\end{subfigure}
	\bigskip
	~ 
	\begin{subfigure}[b]{0.4\textwidth}
		\includegraphics[width=\textwidth]{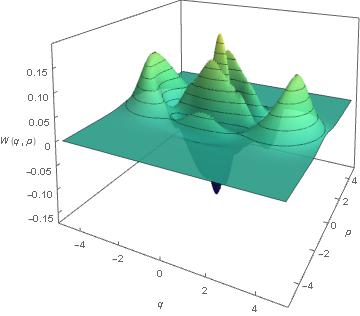}
		\caption{$\vert z\vert=2$, $k_2=1.604011$}
		\label{fig:Wigk2SAO_k_2_0c}
	\end{subfigure}
	\hspace{1cm}
	~ 
	\begin{subfigure}[b]{0.4\textwidth}
		\includegraphics[width=\textwidth]{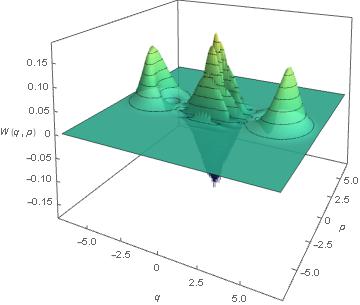}
		\caption{$\vert z\vert=3$, $k_2=1.43425$}
		\label{fig:Wigk2SAO_k_2_0d}
	\end{subfigure}
	\caption{\label{fig:Wigk2SAO_k_2_0}Mandel's $Q_+$-parameter (a) and Wigner function $W_{Z}^+(q,p)$ (b-d) for the states in Eq.~(\ref{64a}) with $a_0=c_2=1$ and different values of $\vert z\vert, \ k_2$. The red line for $Q_+$ marks the values of $k_2$ for which a Poissonian statistics is observed.}
\end{figure}

\noindent while the Wigner functions involving the two scalar states of Eqs.~(\ref{3.20c}, \ref{3.20d}) turn out to be
\begin{subequations}\label{3.37}
	\begin{align}
		&W^{I}_{\alpha,\beta}(q,p)=\frac{1}{\pi}\int_{-\infty}^{\infty}\psi'^{\ast}_{\alpha}(q+y)\psi_{\beta}(q-y)\exp\left(2ipy\right)dy=\sqrt{2}\left[q+ip-\frac{\beta}{\sqrt{2}}\right]W_{\alpha,\beta}(q,p), \label{3.37b} \\
		&W^{II}_{\alpha,\beta}(q,p)=\frac{1}{\pi}\int_{-\infty}^{\infty}\psi^{\ast}_{\alpha}(q+y)\psi'_{\beta}(q-y)\exp\left(2ipy\right)dy=\sqrt{2}\left[q-ip-\frac{\alpha^\ast}{\sqrt{2}}\right]W_{\alpha,\beta}(q,p). \label{3.37c}
	\end{align}
\end{subequations}

\begin{figure}[h!]
	\centering
	\begin{subfigure}[b]{0.4\textwidth}
		\includegraphics[width=\textwidth]{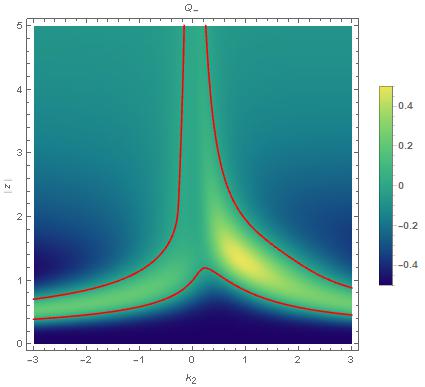}
		\caption{$z\in\mathbb{R}$}
		\label{fig:Wigk2SAO_k_2_1a}
	\end{subfigure}
	\hspace{1cm}
	~ 
	\begin{subfigure}[b]{0.4\textwidth}
		\includegraphics[width=\textwidth]{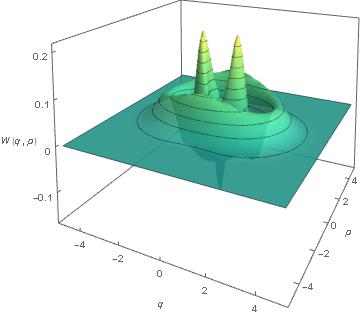}
		\caption{$\vert z\vert=1$, $k_2=2.586$}
		\label{fig:Wigk2SAO_k_2_1b}
	\end{subfigure}
	\bigskip
	~ 
	\begin{subfigure}[b]{0.4\textwidth}
		\includegraphics[width=\textwidth]{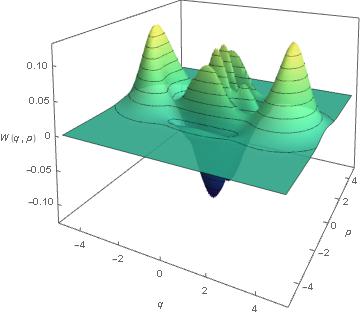}
		\caption{$\vert z\vert=2$, $k_2=0.951075$}
		\label{fig:Wigk2SAO_k_2_1c}
	\end{subfigure}
	\hspace{1cm}
	~ 
	\begin{subfigure}[b]{0.4\textwidth}
		\includegraphics[width=\textwidth]{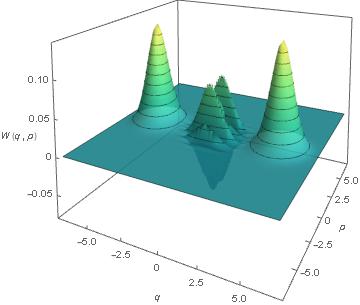}
		\caption{$\vert z\vert=3$, $k_2=0.48326$}
		\label{fig:Wigk2SAO_k_2_1d}
	\end{subfigure}
	\caption{\label{fig:Wigk2SAO_k_2_1}Mandel's $Q_-$-parameter (a) and Wigner function $W_{Z}^-(q,p)$ (b-d) for the states in Eq.~(\ref{64b}) with $a_1=c_1=1$ and different values of $\vert z\vert, \ k_2$. The red line for $Q_-$ marks the values of $k_2$ for which a Poissonian statistics is observed.}
\end{figure}

\subsubsection{Non-classicality criteria for $m=1$.}
By considering first the case with $m=1$, $j=0$ we obtain the Mandel's parameter $Q_Z$ given in Eq.~(\ref{3.38a}) in the Appendix, while the Wigner function $W_Z(q,p)$ turns out to be:
	\begin{equation}
		W_{Z}(q,p)=\mathcal{N}^2\left[\left(\vert a_0\vert^2+\vert c_1\vert^2\right)W_{z}(q,p)+\vert c_1\vert^2k_2^2W'_{z}(q,p)-k_2\left(a_0^\ast c_1W^{II}_{z}(q,p)+a_0 c_1^\ast W^{I}_{z}(q,p)\right)\right]. \label{3.38b}
	\end{equation}
The Mandel's parameter $Q_Z$ as function of $k_2$ is shown in Figure \ref{fig:Wigk2SAO_k_11a}. Although the supercoherent states $\vert Z\rangle$ exhibit in general either sub-Poissonian or super-Poissonian statistics, nonetheless there are $k_2$-values for which $Q_Z=0$ (red line), {\it i.e.}, for the conditions defined by this line they can be considered as semi-classical states.

On the other hand, the corresponding Wigner function $W_Z(q,p)$ acquires negative values for $k_2\neq 0$, which is a sign of an intrinsically quantum nature for these supercoherent states \cite{kenfack04}; however, these negative values tend to disappear as $k_2\rightarrow 0$, \textit{i.e.}, the parameter $k_2$ affects substantially the quantum nature of the states $\vert Z\rangle$ (see Figures \ref{fig:Wigk2SAO_k_11b}-\ref{fig:Wigk2SAO_k_11d}). In particular, for $k_2=0$ it is recovered the Wigner function for the SCS, up to constant factor.  

\begin{figure}[h!]
	\centering
	\begin{subfigure}[b]{0.4\textwidth}
		\includegraphics[width=\textwidth]{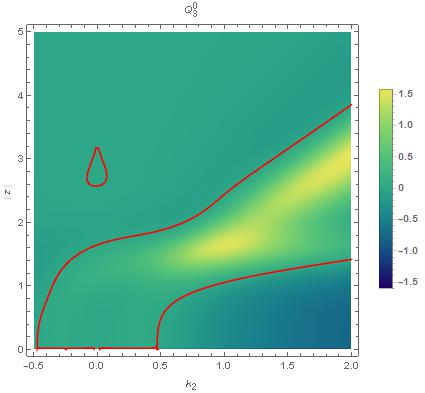}
		\caption{$z\in\mathbb{R}$}
		\label{fig:Wigk2SAO_k_3_0a}
	\end{subfigure}
	\hspace{1cm}
	~ 
	\begin{subfigure}[b]{0.4\textwidth}
		\includegraphics[width=\textwidth]{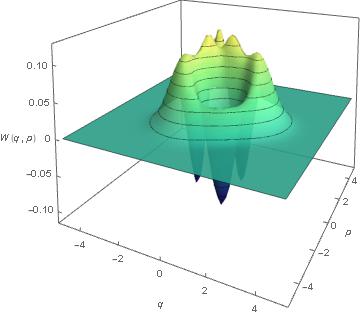}
		\caption{$\vert z\vert=1$, $k_2=-0.351633$}
		\label{fig:Wigk2SAO_k_3_0b}
	\end{subfigure}
	\bigskip
	~ 
	\begin{subfigure}[b]{0.4\textwidth}
		\includegraphics[width=\textwidth]{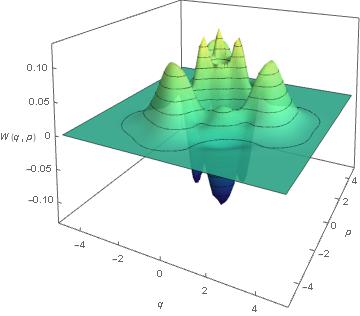}
		\caption{$\vert z\vert=2$, $k_2=0.6805165$}
		\label{fig:Wigk2SAO_k_3_0c}
	\end{subfigure}
	\hspace{1cm}
	~ 
	\begin{subfigure}[b]{0.4\textwidth}
		\includegraphics[width=\textwidth]{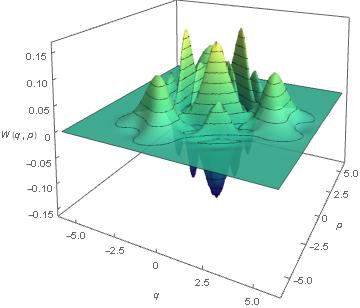}
		\caption{$\vert z\vert=3$, $k_2=1.386432$}
		\label{fig:Wigk2SAO_k_3_0d}
	\end{subfigure}
	\caption{\label{fig:Wigk2SAO_k_3_0}Mandel's $Q_3^0$-parameter (a) and Wigner function $W_{Z,3}^0(q,p)$ (b-d) for the states in Eq.~(\ref{68a}) with $a_0=c_3=1$ and different values of $\vert z\vert, \ k_2$. The red line for $Q_3^0$ marks the values of $k_2$ for which a Poissonian statistics is observed.}
\end{figure}

\subsubsection{Non-classicality criteria for $m=2$.}

\begin{figure}[h!]
	\centering
	\begin{subfigure}[b]{0.4\textwidth}
		\includegraphics[width=\textwidth]{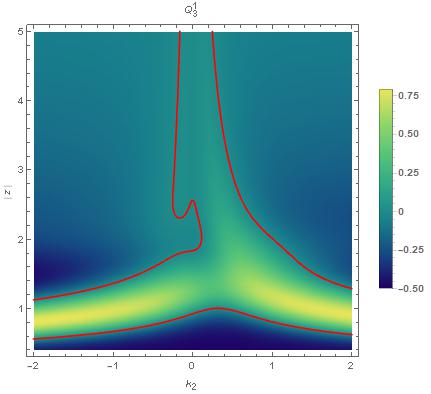}
		\caption{$z\in\mathbb{R}$}
		\label{fig:Wigk2SAO_k_3_1a}
	\end{subfigure}
	\hspace{1cm}
	~ 
	\begin{subfigure}[b]{0.4\textwidth}
		\includegraphics[width=\textwidth]{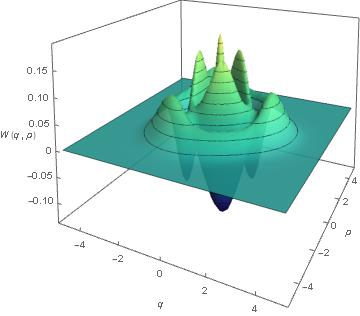}
		\caption{$\vert z\vert=1$, $k_2=-2.94005$}
		\label{fig:Wigk2SAO_k_3_1b}
	\end{subfigure}
	\bigskip
	~ 
	\begin{subfigure}[b]{0.4\textwidth}
		\includegraphics[width=\textwidth]{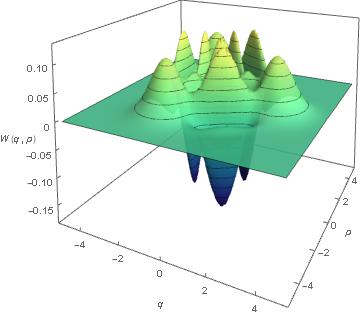}
		\caption{$\vert z\vert=2$, $k_2=0.111063$}
		\label{fig:Wigk2SAO_k_3_1c}
	\end{subfigure}
	\hspace{1cm}
	~ 
	\begin{subfigure}[b]{0.4\textwidth}
		\includegraphics[width=\textwidth]{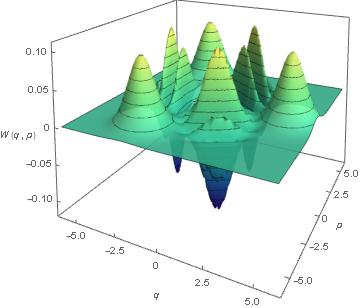}
		\caption{$\vert z\vert=3$, $k_2=0.48317$}
		\label{fig:Wigk2SAO_k_3_1d}
	\end{subfigure}
	\caption{\label{fig:Wigk2SAO_k_3_1}Mandel's $Q_3^1$-parameter (a) and Wigner function $W_{Z,3}^1(q,p)$ (b-d) for the states in Eq.~(\ref{68b}) with $a_1=c_1=1$ and different values of $\vert z\vert, \ k_2$. The red line for $Q_3^1$ marks the values of $k_2$ for which a Poissonian statistics is observed.}
\end{figure}

If we take now $m=2$, $j=0,1$, the Mandel's parameters $Q_+$ and $Q_-$ of equations (\ref{3.40a}) and (\ref{3.40aa}) in the Appendix are obtained, while the Wigner functions $W^\pm_{Z}(q,p)$ turn out to be:
\begin{align}
\nonumber W^\pm_{Z}(q,p)&=\mathcal{N}_\pm^2\Big[\vert\tilde{a}_{2_j}\vert^2W^\pm_{z}(q,p)+\vert \tilde{c}_{2_j}\vert^2W^{\mp}_{z}(q,p)+\vert\tilde{c}_{2_j}\vert^2k_2^2W'^\pm_{z}(q,p)\\
&\quad-k_2\left(\tilde{a}_{2_j}^\ast\tilde{c}_{2_j}W^{II\pm}_{z}(q,p)+\tilde{a}_{2_j}\tilde{c}_{2_j}^\ast W^{I\pm}_{z}(q,p)\right)\Big]. \label{3.40b}
\end{align}

\begin{figure}[h!]
	\centering
	\begin{subfigure}[b]{0.4\textwidth}
		\includegraphics[width=\textwidth]{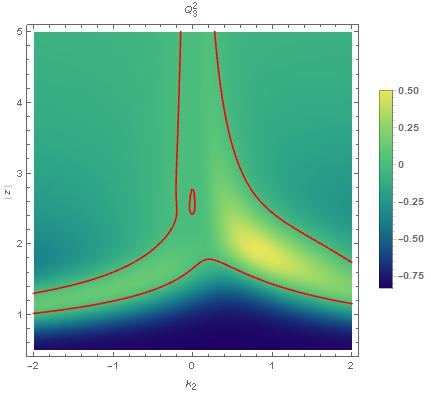}
		\caption{$z\in\mathbb{R}$}
		\label{fig:Wigk2SAO_k_3_2a}
	\end{subfigure}
	\hspace{1cm}
	~ 
	\begin{subfigure}[b]{0.4\textwidth}
		\includegraphics[width=\textwidth]{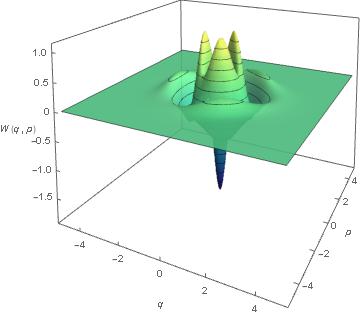}
		\caption{$\vert z\vert=1$, $k_2=-2.116$}
		\label{fig:Wigk2SAO_k_3_2b}
	\end{subfigure}
	\bigskip
	~ 
	\begin{subfigure}[b]{0.4\textwidth}
		\includegraphics[width=\textwidth]{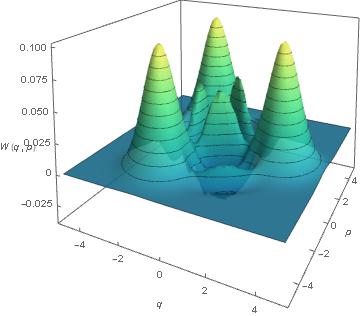}
		\caption{$\vert z\vert=2$, $k_2=-0.419157$}
		\label{fig:Wigk2SAO_k_3_2c}
	\end{subfigure}
	\hspace{1cm}
	~ 
	\begin{subfigure}[b]{0.4\textwidth}
		\includegraphics[width=\textwidth]{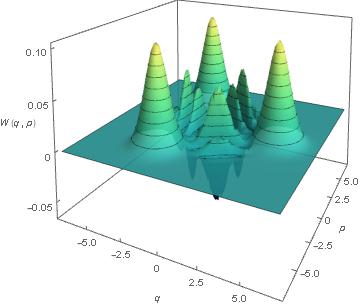}
		\caption{$\vert z\vert=3$, $k_2=-0.206622$}
		\label{fig:Wigk2SAO_k_3_2d}
	\end{subfigure}
	\caption{\label{fig:Wigk2SAO_k_3_2}Mandel's $Q_3^2$-parameter (a) and Wigner function $W_{Z,3}^2(q,p)$ (b-d) for the states in Eq.~(\ref{68c}) with $a_2=c_2=1$ and different values of $\vert z\vert, \ k_2$. The red line for $Q_3^2$ marks the values of $k_2$ for which a Poissonian statistics is observed.}
\end{figure}

The Mandel's $Q_\pm$-parameter as function of $k_2$ for the states $\vert Z\rangle_\pm$ is represented in Figures \ref{fig:Wigk2SAO_k_2_0a} and \ref{fig:Wigk2SAO_k_2_1a}. Once again, these multiphoton supercoherent states exhibit sub-Poissonian and super-Poissonian statistics, but there are also  $k_2$-values for which $Q_\pm=0$ (red line), meaning that on this line these states have a semi-classical behavior.

On the other hand, from Figures \ref{fig:Wigk2SAO_k_2_0} and \ref{fig:Wigk2SAO_k_2_1} it is seen that the Wigner function $W_\pm(q,p)$ for each multiphoton supercoherent state $\vert Z\rangle_\pm$ behaves qualitatively in the same way as its scalar counterpart, showing two localized Gaussian distributions which however interfere with each other, according to the value taken by $k_2$ (see Figures \ref{fig:Wigk2SAO_k_2_0c}, \ref{fig:Wigk2SAO_k_2_0d}, \ref{fig:Wigk2SAO_k_2_1c}, \ref{fig:Wigk2SAO_k_2_1d}) \cite{kenfack04}.

\subsubsection{Non-classicality criteria for $m=3$.}
Finally, if we take $m=3$, $j=0,1,2,$ the corresponding Mandel's parameter $Q_3^j$ of Eq.~(\ref{3.44a}) in the Appendix is obtained, while the Wigner functions $W^j_{Z,3}(q,p)$ turns out to be:
	\begin{align}
\nonumber		W^j_{Z,3}(q,p)&=\mathcal{N}_j^2\Big[\vert \tilde{a}_{3_j}\vert^2W^j_{z,3}(q,p)+\vert \tilde{c}_{3_j}\vert^2W^{3_j-1}_{z,3}(q,p)+\vert \tilde{c}_{3_j}\vert^2k_2^2W'^j_{z,3}(q,p)\\
		&\quad-k_2\left(\tilde{a}_{3_j}^\ast\tilde{c}_{3_j}W^{jII}_{z,3}(q,p)+\tilde{a}_{3_j}\tilde{c}_{3_j}^\ast W^{jI}_{z,3}(q,p)\right)\Big]. \label{3.44b}
	\end{align}
The Mandel's $Q_3^j$-parameter as function of $k_2$ for the states $\vert Z;3,j\rangle$ shows sub-Poissonian and super-Poissonian statistics. Moreover, for some particular values of the parameter $k_2$ it also arises a Poissonian behavior (see red line in Figures \ref{fig:Wigk2SAO_k_3_0a}, \ref{fig:Wigk2SAO_k_3_1a} and \ref{fig:Wigk2SAO_k_3_2a}). As mentioned previously, these multiphoton supercoherent states thus could be considered as semi-classical for these $k_2$-values. 

Concerning the Wigner functions $W_3^j(q,p)$ associated to the states $\vert Z;3,j\rangle$, they behave similarly as their scalar counterparts of Eqs.~(\ref{2.68a}-\ref{2.68c}), with three localized Gaussian distributions interfering to each other according to the value taken by $k_2$ (see Figures \ref{fig:Wigk2SAO_k_3_0c}, \ref{fig:Wigk2SAO_k_3_0d}, \ref{fig:Wigk2SAO_k_3_1c}, \ref{fig:Wigk2SAO_k_3_1d}, \ref{fig:Wigk2SAO_k_3_2c}, \ref{fig:Wigk2SAO_k_3_2d}).

\subsection{Evolution loop and geometric phase}\label{secc3.2.3} 
Let us apply now the evolution operator $\hat{U}(t)=\exp(-i\hat{H}t)$ to the states $\vert Z;m,j\rangle$, which are expressed as a linear combination of the eigenstates of $\hat{H}_{\text{SUSY}}$ (see Eq.~(\ref{3.3b})). We find that
\small
\begin{align}\label{3.52}
	\nonumber \hat{U}(t)\vert Z;m,j\rangle&=\sum_{n=0}^{\infty}\left[\frac{\tilde{a}_{m_j}\tilde{\alpha}^n}{\sqrt{(mn+j)}!}-\frac{(mn+j)\,k_2\tilde{c}_{m_j}\tilde{\alpha}^{n-1}}{\sqrt{(mn+j)!}}\right]\exp\left(-i\omega(mn+j)t\right)\left(
	\begin{array}{c}
		\vert mn+j\rangle \\
		0 \\
	\end{array}
	\right) \\
	\nonumber  &\quad+\sum_{n=0}^{\infty}\frac{\tilde{c}_{m_j}\tilde{\alpha}^{n}\exp\left(-i\omega(mn+m_j)t\right)}{\sqrt{(mn+m_j-1)!}}\left(
	\begin{array}{c}
		0 \\
		|mn+m_j-1\rangle \\
	\end{array}
	\right) \\
	\nonumber&\quad=\exp\left(-i\omega jt\right)\left[\tilde{a}_{m_j}\left(
	\begin{array}{c}
		\vert \tilde{z}(t);m,j\rangle \\
		0 \\
	\end{array}
	\right)+\tilde{c}_{m_j}\left(
	\begin{array}{c}
		-k_2\exp\left(-i\omega mt\right)\vert(\tilde{z}(t))';m,j\rangle \\
		\exp\left(-i\omega(m_j-j)t\right)|\tilde{z}(t);m,m_j-1\rangle \\
	\end{array}
	\right)\right]\\
	&\quad\neq \exp\left(-i\omega jt\right)\vert Z(t);m,j\rangle.
\end{align}
where $\tilde{z}(t)=\tilde{\alpha}\exp(-i\omega mt)=[z\exp(-i\omega t)]^m=[z(t)]^m$.
\normalsize

\begin{figure}[h!]
	\centering
	\begin{subfigure}[b]{0.34\textwidth}
		\includegraphics[width=\textwidth]{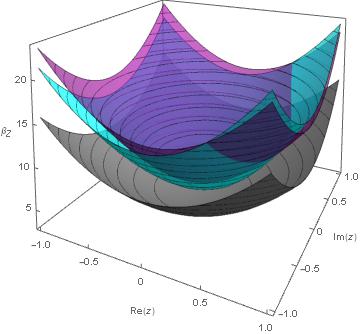}
		\caption{$\beta_Z\equiv\beta_1^0$}
		\label{fig:geo_phase_a}
	\end{subfigure}
	\hspace{1cm}
	~ 
	\begin{subfigure}[b]{0.34\textwidth}
		\includegraphics[width=\textwidth]{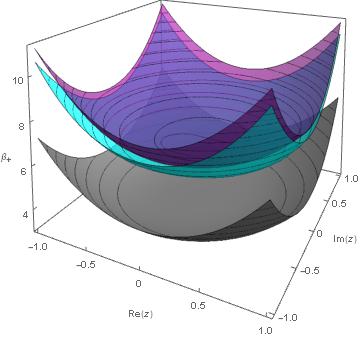}
		\caption{$\beta_+\equiv\beta_2^0$}
		\label{fig:geo_phase_b}
	\end{subfigure}
	\bigskip
	~ 
	\begin{subfigure}[b]{0.34\textwidth}
		\includegraphics[width=\textwidth]{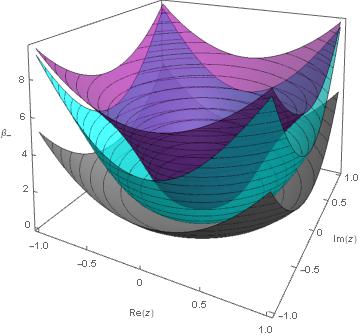}
		\caption{$\beta_-\equiv\beta_2^1$}
		\label{fig:geo_phase_c}
	\end{subfigure}
	\hspace{1cm}
	~ 
	\begin{subfigure}[b]{0.34\textwidth}
		\includegraphics[width=\textwidth]{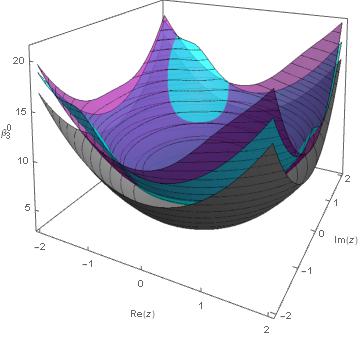}
		\caption{$\beta_3^0$}
		\label{fig:geo_phase_d}
	\end{subfigure}
	\bigskip
	~ 
	\begin{subfigure}[b]{0.34\textwidth}
		\includegraphics[width=\textwidth]{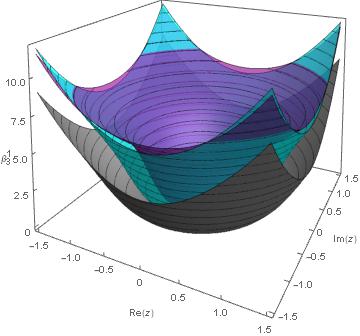}
		\caption{$\beta_3^1$}
		\label{fig:geo_phase_e}
	\end{subfigure}
	\hspace{1cm}
	~ 
	\begin{subfigure}[b]{0.34\textwidth}
		\includegraphics[width=\textwidth]{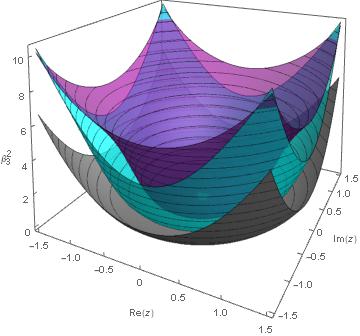}
		\caption{$\beta_3^2$}
		\label{fig:geo_phase_f}
	\end{subfigure}
	\caption{\label{fig:geo_phase_MCS}Geometric phase $\beta_m^j$ for the states $\vert Z;m,j\rangle$, $m=1,2,3$ with different values of the parameter $k_2$: $k_2=-4$ (purple), $k_2=0$ (gray) and $k_2=2$ (cyan). The case with $k_2=0$ produces the allowed minimum values for $\beta_m^j$.}
\end{figure}

Although the scalar multiphoton coherent states evolve coherently, so that $\hat{U}(t)$ transforms any of these states into another of the same family, Eq.~(\ref{3.52}) indicates, however, that for the multiphoton supercoherent states in general this is no longer true. In fact, only for $\tilde{c}_{m_j}=0$ such a property is satisfied. Despite, the multiphoton supercoherent states turn out to be cyclic, since by taking $t\equiv\tau=2\pi/\omega m$ in Eq.~(\ref{3.52}) it is obtained:
\begin{equation}\label{3.53}
	\hat{U}(\tau)\vert Z;m,j\rangle=\exp\left(i\phi\right)\vert Z;m,j\rangle, \qquad  \phi=-2\pi\frac{j}{m}.
\end{equation}

Eq.~(\ref{3.53}) characterizes an important property that the states $\vert Z;m,j\rangle$ share with the multiphoton coherent states of Eq.~(\ref{2.28}): both states recover their initial condition after a time interval $\tau=\tau_{\text{classical}}/m$, where $\tau_{\text{classical}}=2\pi/\omega$. This fact does not have any classical counterpart, except for $m=1$ where both periods coincide.

Finally, from Eq.~(\ref{2.93}) the geometric phase $\beta_m^j$ associated to the multiphoton supercoherent states in each subspace $\mathcal{H}_{j}$ can be calculated, leading to (see also Figure \ref{fig:geo_phase_MCS}):
\begin{equation}\label{3.61}
	\beta_m^j=-2\pi\frac{j}{m}+\frac{2\pi}{m}\omega^{-1}\frac{\langle Z;m,j\vert\hat{H}_{\text{SUSY}}\vert Z;m,j\rangle}{\langle Z;m,j\vert Z;m,j\rangle}, \quad j=0,1,2,\dots,m-1,
\end{equation}
where the mean energy value is given by
\begin{align}\label{3.54}
	&\nonumber \omega^{-1}\langle Z;m,j\vert\hat{H}_{\text{SUSY}}\vert Z;m,j\rangle= \vert\tilde{a}_{m_j}\vert^2\langle z;m,j\vert\hat{N}\vert z;m,j\rangle+\vert\tilde{c}_{m_j}\vert^2\langle z;m,m_j-1\vert\hat{N}+\hat{1}\vert z;m,m_j-1\rangle\\
	&\qquad+\vert\tilde{c}_{m_j}\vert^2k_2^2\langle z';m,j\vert\hat{N}\vert z';m,j\rangle-2k_2\text{Re}(\tilde{a}_{m_j}\tilde{c}^\ast_{m_j}\langle z';m,j\vert\hat{N}\vert z;m,j\rangle).
\end{align}

In particular, if we choose $m=1$ and $k_2=0$ in Eq.~(\ref{3.14}), we will obtain the simplest supercoherent states, which share many properties with the SCS \cite{aragone,hussin}, as the Heisenberg uncertainty relation $(\sigma_q)_Z(\sigma_p)_Z$, the Mandel's $Q$-parameter, the Wigner function $W_Z(q,p)$, and the evolution loop period $\tau=2\pi/\omega$.

\section{Conclusions}\label{sec4}
Starting from the harmonic oscillator, through a clever choice of generalized annihilation and creation operators, we have seen that it is possible to define different algebraic structures, as the Heisenberg-Weyl algebras (HWA) or the polynomial Heisenberg algebras (PHA). The PHA are generalizations of the HWA which appear when replacing the standard annihilation and creation operators $\hat{a}$, $\hat{a}^\dagger$ by $m$-th order differential ones $\hat{\mathcal{L}}^-_m$, $\hat{\mathcal{L}}^\dagger_m$ \cite{fh99,fhn94,fno95,carballo04,bermudez14,celeita16}.

In particular, if we take $\hat{\mathcal{L}}^-_m\equiv\hat{a}^m$, $\hat{\mathcal{L}}^\dagger_m\equiv\hat{a}^{\dagger m}$, then the set of operators $\{\hat{H},\hat{a}^m,\hat{a}^{\dagger m}\}$ generates a PHA, which allows to construct $m$ infinite energy ladders for the harmonic oscillator Hamiltonian $H$. In each ladder there is an extremal state $\vert \psi^j_{0}\rangle = \vert j\rangle$ associated to a minimum energy level $E^j_0= E_j=j+1/2$, $j=0,1,\dots,m-1$. As a consequence, the Hilbert space $\mathcal{H}$ decomposes as the direct sum of $m$ orthogonal subspaces $\mathcal{H}_{j}$, whose basis vectors are constructed by applying iteratively the operator $\hat{a}^{\dagger m}$ onto the extremal states $\vert \psi^j_0\rangle$.

Once the algebraic structure has been characterized, it is straightforward to construct the so-called multiphoton coherent states (MCS) as eigenstates of the generalized annihilation operator $\hat{a}^m$ \cite{cdf18,perelomov72,barut71,buzek90,buzek901,sun92,jex93}. It has been shown that these states can be expressed either as linear combinations of the energy eigenstates of each subspace $\mathcal{H}_{j}$ or as superpositions of standard coherent states (SCS) \cite{schrodinger35,gagen95}, the last ones being obtained as well as particular cases of the MCS with $m=1$. The SCS are minimum uncertainty states, but in general this is no longer true for the MCS, mainly due to the fact that the extremal state contributing to each MCS is not the ground state of the oscillator anymore. Besides, according to the behavior of the Mandel's $Q$-parameter and the Wigner function, the multiphoton coherent states become intrinsically quantum states for $m>1$ \cite{kenfack04}. Finally, since the energy spectrum for the harmonic oscillator is equidistant, a partial evolution loop on each subspace $\mathcal{H}_{j}$ is produced, whose period turns out to be the fraction $1/m$ of the classical period. As a consequence, the MCS are cyclic states, with the same period as the partial loop, and the corresponding geometric phase has been explicitly calculated.

On the other hand, the supercoherent states for the SUSY harmonic oscillator turn out to be expressed in terms of the corresponding coherent states of the scalar case \cite{aragone,zypman,hussin,erik16}. As a consequence, the multiphoton supercoherent states for the supersymmetric harmonic oscillator were also explored trying to see if the (scalar) multiphoton coherent states were also involved in such states. In order to do that, we considered a particular form of the supersymmetric annihilation operator $\hat{A}_{\text{SUSY}}$ and built then its $m$-th power, where $k_2$ was left as a free real parameter (see Eq.~(\ref{3.14})). This choice allowed us to analyze the effect of the parameter $k_2$ onto the quantum nature of the multiphoton supercoherent states, for at least three particular values of $m$.

For the simplest case, with $m=1$ and $k_2=0$, the Heisenberg uncertainty relation and Wigner function were qualitatively the same as the corresponding results for the standard coherent states. Moreover, the Mandel's $Q$-parameter vanishes ($Q_Z=0$) for $k_2=0$ (see also Figure \ref{fig:Wigk2SAO_k_11}a). On the other hand, for the multiphoton supercoherent states with $m=2,3$, the minimum value of the uncertainty product $(\sigma_q)_Z(\sigma_p)_Z$ (which arises in the limit $\alpha\rightarrow0$) changes as $\vert k_2\vert$ does (see Figures \ref{fig:Inck2SAO_k_2_0} and \ref{fig:Inck2SAO_k_3_0}). Meanwhile, the states $\vert Z;m,j\rangle$ exhibit sub-Poissonian statistics for any $m$, meaning that these are in general non-classical states, as the corresponding Wigner functions also show for some values of $k_2$ (see Figures \ref{fig:Wigk2SAO_k_11}-\ref{fig:Wigk2SAO_k_3_2}). However, there are some values of $k_2$ for which $Q=0$ (red line in Figures \ref{fig:Wigk2SAO_k_11}a-\ref{fig:Wigk2SAO_k_3_2}a), which could be interpreted as the existence of a particular domain of $k_2$ for which the non-classical effects in the multiphoton supercoherent states disappear.

In addition, by taking into account that the SUSY harmonic oscillator has an equidistant spectrum, it has been possible to study both, the evolution loop of the system as well as the geometric phases $\beta_m^j$ associated to the multiphoton supercoherent states. Since both, the SCS and the MCS in the scalar case are cyclic states recovering their initial condition after the time interval $\tau=\tau_{\text{classical}}/m$, then for $m>1$ they can be considered as states without any classical counterpart. Concerning the multiphoton supercoherent states, the same situation appears for $m=1$ and $k_2=0$. Moreover, as Eq.~(\ref{3.61}) shows, the geometric phase $\beta_m^j$ for the states $\vert Z;m,j\rangle$ is similar to the one of its scalar counterpart (see Eq.~(\ref{2.105b})), having to subtract just the ground state energy. On the other hand, the case with $k_2=0$ defines a lower bound for the geometric phase associated to the multiphoton supercoherent states considered in this paper (see Figure \ref{fig:geo_phase_MCS}).

It is important to remark that the form chosen for the supersymmetric annihilation operator allowed us to construct and describe in a relatively simple way the multiphoton supercoherent states, as well as to study the effect of the parameters $k_2$ onto their intrinsically quantum nature. It would be interesting to consider the most general case, {\it i.e.}, to allow arbitrary values for the four parameters $k_i$, and to obtain the extended family of eigenstates of the SAO, which presumably will have a richer and more general structure.

Let us conclude by noticing that the SUSY harmonic oscillator can be used as a toy model in the description of the matter-radiation interaction. In such a case the multiphoton supercoherent states could help to understand and/or explain the simplest physical processes involving the simultaneous absortion or emission of several photons \cite{dsi06}. Of course, this description would be valid just in the limit where the matter-radiation interaction can be neglected and in the resonant regime, since in this case the SUSY harmonic oscillator Hamiltonian is recovered from the more realistic description supplied by the well known Jaynes-Cumming model \cite{dh02}, which along the years has proved so successful for describing this kind of phenomena \cite{dsi06}.

\subsubsection*{Acknowledgments}
The authors acknowledge the support of Conacyt, as well as the comments and suggestions of the referees of this paper. EDB also acknowledges the warm hospitality at Department of Theoretical Physics of the University of Valladolid.

\appendix

\section*{Appendix}\label{sec5}
\renewcommand{\theequation}{A.\arabic{equation}}
\setcounter{equation}{0}

Uncertainty square for the operator $\hat{s}$ of Eq.~(\ref{3.1.58}) in the multiphoton supercoherent states with $m=1, \ j=0$ (Eq.~(\ref{sm1j0})), with $m=2, \ j=0,1$ (Eqs.~(\ref{3.28a}, \ref{3.28b})) and with $m=3, \ j=0,1,2$ (Eq.~(\ref{3.31})):
\small
\begin{subequations}
	\begin{align}\label{sm1j0}
		\nonumber(\sigma_s)^2_Z&=\mathcal{N}^2\frac{\exp(\vert z\vert^2)}{2}\left[(\vert a_0\vert^2+\vert c_1\vert^2)(2\vert z\vert^2+1+2(-1)^k([\text{Re}(z)]^2-[\text{Im}(z)]^2))\right. \\
		\nonumber&\quad+\vert c_1\vert^2k^2_2(2\vert z\vert^2+1+2(-1)^k([\text{Re}(z)]^2-[\text{Im}(z)]^2))(\vert z\vert^2+3) \\
		\nonumber&\quad-2k_2\mathrm{Re}[a_0c_1^\ast[(2\vert z\vert^2+3)z+(-1)^k((\vert z\vert^2+2)z^\ast+z^3)]] \\
		\nonumber&\quad+(-1)^{k+1}\mathcal{N}^4\frac{\exp(2\vert z\vert^2)}{2}\left[(\vert a_0\vert^2+\vert c_1\vert^2)(z+(-1)^kz^\ast)+\vert c_1\vert^2k_2^2(\vert z\vert^2+2)(z+(-1)^kz^\ast)\right. \\
		&\quad\left.-k_2(a_0^\ast c_1(\vert z\vert^2+1+(-1)^kz^{\ast 2})+a_0c_1^\ast(z^2+(-1)^k(\vert z\vert^2+1)))\right]^2,
		\\
		\nonumber(\sigma_s)^2_{Z+}&=\mathcal{N}_+^2\frac{\cosh(\vert z\vert^2)}{2}\left[\vert\tilde{a}_{2_0}\vert^2(2\vert z\vert^2\tanh(\vert z\vert^2)+1+2(-1)^k([\text{Re}(z)]^2-[\text{Im}(z)]^2))\right.\\
		\nonumber&+\vert\tilde{c}_{2_0}\vert^2(2\vert z\vert^2+\tanh(\vert z\vert^2)+2(-1)^k\tanh(\vert z\vert^2)([\text{Re}(z)]^2-[\text{Im}(z)]^2))+\vert\tilde{c}_{2_0}\vert^2k_2^2\times \\
		\nonumber&\times\left(2\vert z\vert^4\tanh(\vert z\vert^2)+7\vert z\vert^2+3\tanh(\vert z\vert^2)+2(-1)^k(\vert z\vert^2+3\tanh(\vert z\vert^2))([\text{Re}(z)]^2-[\text{Im}(z)]^2)\right) \\
		&\left.-k_2\left(\left(2\vert z\vert^2+3\tanh(\vert z\vert^2)+(-1)^k(\vert z\vert^2\tanh(\vert z\vert^2)+2)\right)\mathrm{Re}[\tilde{a}_{2_0}^\ast\tilde{c}_{2_0}z]+\tanh(\vert z\vert^2) \mathrm{Re}[\tilde{a}_{2_0}^\ast\tilde{c}_{2_0}z^{\ast 3}]\right)\right], \label{3.28a} \\
		\nonumber(\sigma_s)^2_{Z-}&=\mathcal{N}_-^2\frac{\sinh(\vert z\vert^2)}{2}\left[\vert\tilde{a}_{2_1}\vert^2(2\vert z\vert^2\coth(\vert z\vert^2)+1+2(-1)^k([\text{Re}(z)]^2-[\text{Im}(z)]^2))\right.\\
		\nonumber&+\vert\tilde{c}_{2_1}\vert^2(2\vert z\vert^2+\coth(\vert z\vert^2)+2(-1)^k\tanh(\vert z\vert^2)([\text{Re}(z)]^2-[\text{Im}(z)]^2))+\vert\tilde{c}_{2_1}\vert^2k_2^2\times \\
		\nonumber&\times\left(2\vert z\vert^4\coth(\vert z\vert^2)+7\vert z\vert^2+3\coth(\vert z\vert^2)+2(-1)^k(\vert z\vert^2+3\coth(\vert z\vert^2))([\text{Re}(z)]^2-[\text{Im}(z)]^2)\right) \\
		&\left.-k_2\left(\left(2\vert z\vert^2+3\coth(\vert z\vert^2)+(-1)^k(\vert z\vert^2\coth(\vert z\vert^2)+2)\right)\mathrm{Re}[\tilde{a}_{2_1}^\ast\tilde{c}_{2_1}z]+\coth(\vert z\vert^2) \mathrm{Re}[\tilde{a}_{2_1}^\ast\tilde{c}_{2_1}z^{\ast 3}]\right)\right], \label{3.28b}
		\\
		\nonumber(\sigma_s)^2_{jZ}&=\frac{\mathcal{N}_j^2}{2}\Big[\vert\tilde{a}_{3_j}\vert^2\left(2\vert z\vert^2[\mathcal{N}_3^{3_j-1}]^{-2}+[\mathcal{N}_3^j]^{-2}\right)+\vert\tilde{c}_{3_j}\vert^2\left(2\vert z\vert^2[\mathcal{N}_3^{s_j}]^{-2}+[\mathcal{N}_3^{3_j-1}]^{-2}\right) \\
\nonumber		&\quad+\vert\tilde{c}_{3_j}\vert^2k_2\left(2\vert z\vert^4[\mathcal{N}_3^j]^{-2}+7\vert z\vert^2[\mathcal{N}_3^{s_j}]^{-2}+3[\mathcal{N}_3^{3_j-1}]^{-2}\right) \\
		&\quad-k_2\left(2\vert z\vert^2[\mathcal{N}_3^{s_j}]^{-2}+3[\mathcal{N}_3^{3_j-1}]^{-2}\right)\mathrm{Re}[\tilde{a}_{3_j}^\ast\tilde{c}_{3_j}z^\ast]\Big], \label{3.31}
	\end{align}
\end{subequations}
\normalsize
where $3_j=3\delta_{0j}+j$ and $s_j=(j+1)(1-\delta_{2j})$. 


Mandel $Q$-parameter for the multiphoton supercoherent states with $m=1, \ j=0$ (Eq.~(\ref{3.38a})), with $m=2, \ j=0,1$ (Eqs.~(\ref{3.40a}, \ref{3.40aa})) and with $m=3, \ j=0,1,2$ (Eq.~(\ref{3.44a})):
\small
\begin{subequations}
	\begin{align}
		\nonumber Q_Z&=\vert z\vert^2\left[\left(\vert a_0\vert^2+\vert c_1\vert^2\right)\vert z\vert^2+\vert c_1\vert^2k^2_2\left((\vert z\vert^2+1)^2+\vert z\vert^2\right)-2k_2\left(\vert z\vert^2+1\right)\text{Re}[a_0^\ast c_1 z^\ast]\right]^{-1}\times \\
		\nonumber&\quad\times\left[\left(\vert a_0\vert^2+\vert c_1\vert^2\right)\vert z\vert^2+\vert c_1\vert^2k^2_2\left((\vert z\vert^2+2)^2+\vert z\vert^2\right)-2k_2\left(\vert z\vert^2+2\right)\text{Re}[a_0^\ast c_1 z^\ast]\right] \\
		&\quad -\mathcal{N}^2\exp(\vert z\vert^2)\left[\left(\vert a_0\vert^2+\vert c_1\vert^2\right)\vert z\vert^2+\vert c_1\vert^2k^2_2\left((\vert z\vert^2+1)^2+\vert z\vert^2\right)-2k_2\left(\vert z\vert^2+1\right)\text{Re}[a_0^\ast c_1 z^\ast]\right], \label{3.38a} 
		\\
		\nonumber Q_+&=\vert z\vert^2\left[\left(\vert \tilde a_{2_0}\vert^2\tanh(\vert z\vert^2)+\vert\tilde{c}_{2_0}\vert^2\right)\vert z\vert^2+\vert\tilde{c}_{2_0}\vert^2k_2^2\left((\vert z\vert^4+1)\tanh(\vert z\vert^2)+3\vert z\vert^2\right)\right.\\
		\nonumber&\quad\left.-2k_2\left(\vert z\vert^2+\tanh(\vert z\vert^2)\right)\text{Re}[\tilde{a}_{2_0}^\ast\tilde{c}_{2_0}z^\ast]\right]^{-1}\left[\left(\vert \tilde {a}_{2_0}\vert^2+\vert\tilde{c}_{2_0}\vert^2\tanh(\vert z\vert^2)\right)\vert z\vert^2\right.\\
		\nonumber&\quad\left.+\vert\tilde{c}_{2_0}\vert^2k_2^2\left(\vert z\vert^4+5\vert z\vert^2\tanh(\vert z\vert^2)+4\right)-2k_2\left(\vert z\vert^2\tanh(\vert z\vert^2)+2\right)\text{Re}[\tilde{a}_{2_0}^\ast\tilde{c}_{2_0}z^\ast]\right] \\
		\nonumber&\quad -\mathcal{N}_+^2\cosh(\vert z\vert^2)\left[\left(\vert \tilde a_{2_0}\vert^2\tanh(\vert z\vert^2)+\vert\tilde{c}_{2_0}\vert^2\right)\vert z\vert^2+\vert\tilde{c}_{2_0}\vert^2k_2^2\left((\vert z\vert^4+1)\tanh(\vert z\vert^2)+3\vert z\vert^2\right)\right.\\
		&\quad\left.-2k_2\left(\vert z\vert^2+\tanh(\vert z\vert^2)\right)\text{Re}[\tilde{a}_{2_0}^\ast\tilde{c}_{2_0}z^\ast]\right],  \label{3.40a}\\
		\nonumber Q_-&=\vert z\vert^2\left[\left(\vert \tilde a_{2_1}\vert^2\coth(\vert z\vert^2)+\vert\tilde{c}_{2_1}\vert^2\right)\vert z\vert^2+\vert\tilde{c}_{2_1}\vert^2k_2^2\left((\vert z\vert^4+1)\coth(\vert z\vert^2)+3\vert z\vert^2\right)\right.\\
		\nonumber&\quad\left.-2k_2\left(\vert z\vert^2+\coth(\vert z\vert^2)\right)\text{Re}[\tilde{a}_{2_1}^\ast\tilde{c}_{2_1}z^\ast]\right]^{-1}\left[\left(\vert \tilde {a}_{2_1}\vert^2+\vert\tilde{c}_{2_1}\vert^2\coth(\vert z\vert^2)\right)\vert z\vert^2\right.\\
		\nonumber&\quad\left.+\vert\tilde{c}_{2_1}\vert^2k_2^2\left(\vert z\vert^4+5\vert z\vert^2\coth(\vert z\vert^2)+4\right)-2k_2\left(\vert z\vert^2\coth(\vert z\vert^2)+2\right)\text{Re}[\tilde{a}_{2_1}^\ast\tilde{c}_{2_1}z^\ast]\right] \\
		\nonumber&\quad -\mathcal{N}_-^2\sinh(\vert z\vert^2)\left[\left(\vert \tilde a_{2_1}\vert^2\coth(\vert z\vert^2)+\vert\tilde{c}_{2_1}\vert^2\right)\vert z\vert^2+\vert\tilde{c}_{2_1}\vert^2k_2^2\left((\vert z\vert^4+1)\coth(\vert z\vert^2)+3\vert z\vert^2\right)\right.\\
		&\quad\left.-2k_2\left(\vert z\vert^2+\coth(\vert z\vert^2)\right)\text{Re}[\tilde{a}_{2_1}^\ast\tilde{c}_{2_1}z^\ast]\right], \label{3.40aa} 
		\\
		\nonumber Q_3^j&=\vert z\vert^2\left[\left(\vert\tilde{a}_{3_j}\vert^2[\mathcal{N}_3^{3_j-1}]^{-2}+\vert\tilde{c}_{3_j}\vert^2[\mathcal{N}_3^{s_j}]^{-2}\right)\vert z\vert^2+\vert\tilde{c}_{3_j}\vert^2k_2^2\left(\vert z\vert^4[\mathcal{N}_3^j]^{-2}+3\vert z\vert^2[\mathcal{N}_3^{s_j}]^{-2}+[\mathcal{N}_3^{3_j-1}]^{-2}\right)\right.\\
		\nonumber&\quad\left.-2k_2\left(\vert z\vert^2[\mathcal{N}_3^{s_j}]^{-2}+[\mathcal{N}_3^{3_j-1}]^{-2}\right)\text{Re}[\tilde{a}_{3_j}^\ast\tilde{c}_{3_j}z^\ast]\right]^{-1}\Big[\left(\vert\tilde{a}_{3_j}\vert^2[\mathcal{N}_3^{s_j}]^{-2}+\vert\tilde{c}_{3_j}\vert^2[\mathcal{N}_3^{j}]^{-2}\right)\vert z\vert^2\\
		\nonumber&\quad+\vert\tilde{c}_{3_j}\vert^2k_2^2\left(\vert z\vert^4[\mathcal{N}_3^{3_j-1}]^{-2}+5\vert z\vert^2[\mathcal{N}_3^{j}]^{-2}+4[\mathcal{N}_3^{s_j}]^{-2}\right)-2k_2\left(\vert z\vert^2[\mathcal{N}_3^{j}]^{-2}+2[\mathcal{N}_3^{s_j}]^{-2}\right)\text{Re}[\tilde{a}_{3_j}^\ast\tilde{c}_{3_j}z^\ast]\Big]\\
		\nonumber&\quad	-\mathcal{N}_j^2\left[\left(\vert\tilde{a}_{3_j}\vert^2[\mathcal{N}_3^{3_j-1}]^{-2}+\vert\tilde{c}_{3_j}\vert^2[\mathcal{N}_3^{s_j}]^{-2}\right)\vert z\vert^2+\vert\tilde{c}_{3_j}\vert^2k_2^2\left(\vert z\vert^4[\mathcal{N}_3^j]^{-2}+3\vert z\vert^2[\mathcal{N}_3^{s_j}]^{-2}+[\mathcal{N}_3^{3_j-1}]^{-2}\right)\right.\\
		&\quad\left.-2k_2\left(\vert z\vert^2[\mathcal{N}_3^{s_j}]^{-2}+[\mathcal{N}_3^{3_j-1}]^{-2}\right)\text{Re}[\tilde{a}_{3_j}^\ast\tilde{c}_{3_j}z^\ast]\right]. \label{3.44a} 
	\end{align}
\end{subequations}
\normalsize

\end{document}